\documentclass[10pt,twocolumn, journal]{IEEEtran}
\usepackage{amsmath,amssymb,amsthm,mathrsfs,amsfonts,dsfont,mathtools}
\usepackage[caption=false]{subfig}
\usepackage[ruled,lined,linesnumbered]{algorithm2e}
\usepackage{setspace}
\SetAlFnt{\small}
\SetAlCapFnt{\small}
\SetAlCapNameFnt{\small}

\usepackage{cite, color}
\usepackage{graphicx,epstopdf}
\usepackage{bmpsize}
\usepackage{hyperref}

\def\eps{\varepsilon}
\makeatletter
\newcommand{\vast}{\bBigg@{2.7}}
\newcommand{\Vast}{\bBigg@{4}}
\makeatother

\newcommand{\argmin}{\operatornamewithlimits{argmin}}
\def\IH{\mathcal{I}_{\mathcal{H}}}
\def\IM{\mathcal{I}_{\mathcal{M}}}
\def\eqd{{\,{\buildrel d \over =}\,}}

\newtheorem{theorem}{Theorem}
\newtheorem{lemma}{Lemma}

\ifCLASSINFOpdf
\else
\fi

\ifCLASSOPTIONcaptionsoff
  \newpage
\fi

\begin{document}


\title{Analysis of Multi-Hop Outdoor $ 60 $ GHz   Wireless Networks with Full-Duplex Buffered Relays }
\author{Guang~Yang, \emph{Student Member, IEEE}, Ming~Xiao, \emph{Senior Member, IEEE}, Hussein~Al-Zubaidy, \emph{Senior Member, IEEE}, Yongming~Huang, \emph{Senior Member, IEEE}, and~James~Gross, \emph{Senior Member, IEEE}
\thanks{This work was supported partly by National Natural Science Foundation of China under Grant 61371105, National 973 Programs 2013CB329001 and EU Marie Curie Project, QUICK, No. 612652.}
\thanks{G. Yang, M. Xiao, J. Gross, and H. Al-Zubaidy are with  the Communication Theory Department, KTH Royal Institute of Technology,
Stockholm, Sweden (Email: \{gy, mingx, hzubaidy\}@kth.se, james.gross@ee.kth.se).}
\thanks{Y. Huang is with the School of Information Science
and Engineering, Southeast University, Nanjing 210096, China (Email:
huangym@seu.edu.cn).}
}

\maketitle

\begin{abstract}
The abundance of unlicensed spectrum in the $ 60 $ GHz band makes it an attractive alternative for future wireless communication systems. Such systems are expected to provide data transmission rates in the order of multi-gigabits per second in order to satisfy the ever-increasing demand for high rate data communication. 
Unfortunately, $ 60 $~GHz radio is subject to severe path loss which limits its usability for long-range outdoor communication. In this work, we propose a  multi-hop $ 60 $~GHz wireless network for outdoor communication where multiple full-duplex buffered relays are used to extend the communication range while providing end-to-end performance guarantees to the traffic traversing the network.
We provide a cumulative service process characterization for the $ 60 $~GHz outdoor propagation channel with self-interference in terms of the moment generating function (MGF) of its channel capacity. We then use this characterization to compute probabilistic upper bounds on the overall network performance, i.e., total backlog and end-to-end delay.  Furthermore, we study the effect of self-interference on the network performance and  propose an optimal power allocation scheme to mitigate its impact in order to enhance network performance. Finally, we investigate the relation between relay density and network performance under a total power budget constraint. We show that increasing relay density may have adverse effects on network performance unless self-interference can be kept sufficiently small.
\end{abstract}


\section{Introduction}
\label{sec:Intro}
With rapidly increasing demands on network service, wireless communications over $ 60 $ GHz spectrum (also often referred to as millimeter wave: mmWave) becomes a promising technology to improve the network throughput for future communication systems \cite{rappaport2013millimeter}. Compared to conventional wireless communications in lower frequency bands, $ 60 $~GHz  wireless communications have significant advantages which include considerably broader bandwidth, lower cost electronics, and higher gain directional antenna implementations \cite{daniels200760}. These attributes make mmWave a promising solution for the wireless backhaul \cite{guo200760}, since the initial cost of fiber optic backhaul tends to be quite high and the conventional microwave based backhaul networks cannot support the throughput requirements of future networks.

The $ 60 $~GHz technology can be utilized for both indoor and outdoor communications.
A significant amount of experimental results for investigating the indoor $ 60 $~GHz wireless personal area networks (WPAN) were reported in  recent years \cite{ moraitis2006measurements, smulders2009statistical, geng2009millimeter, reig2014fading, yang2015maximum}. 
On the other hand, very few studies are devoted to study the outdoor $ 60 $~GHz channels.
In \cite{ben2011millimeter}, a channel sounder is deployed to estimate the outdoor $60$~GHz channel using a $ 59 $~GHz horn antenna. Results show that the path loss exponent for the $60$~GHz channel is between $ 2 $ and $ 2.5 $ for the outdoor environment, such as airport fields, urban streets or tunnels. In \cite{violette1988millimeter}, antennas with narrow beamwidth were used to measure the path loss in urban street environments for the line-of-sight (LOS) and non-line-of-sight (NLOS) scenarios. 
Also, the small-scale fading effects were shown to be negligible in the $ 60 $~GHz band due to the short wavelength, i.e., $ \lambda=5 $ mm,  \cite{williamson1997investigating, geng2009millimeter}. Therefore, the channel fading is dominated by the shadowing effect which is generally modeled as a log-normal random variable.

In light of the above, it is clear that the use of the $ 60 $ GHz band is limited to short distances LOS communications, usually below 500 meters. 
In order to overcome larger distances or obstructed paths, a strategically placed store-and-forward relay node may be used to form a multi-hop 60~GHz wireless network. Thus it is our claim that multi-hop communication can be utilized to mitigate the effects of path loss over long distances and/or the effect of NLOS while maintaining the traffic flows' quality of service (QoS) requirements. However, in this case an understanding of corresponding network performance in terms of end-to-end delay and loss probability becomes key to support real-time mission and critical applications, e.g., online banking, remote health, transportation systems operation and control, and electric power systems. Nevertheless, an analytical model for the multi-hop 60~GHz outdoor network does not exist and its performance is not yet understood. 

In this work we provide a probabilistic end-to-end delay and backlog analysis of such networks in terms of the underlying channel parameters. This analysis can be used as a guideline for planning and operating QoS-driven multi-hop 60~GHz network.  
The analysis of multi-hop 60~GHz wireless networks poses two main challenges: (i) the service process characterization for the 60~GHz fading channel, and (ii) multi-hop network performance analysis. The first challenge comes from the random nature of the wireless 60~GHz fading channel which results in time varying channel capacity, the second challenge is a direct result of the limitations and strict assumptions of the traditional queuing theory, which is the main tool for network analysis, when applied to queuing networks. 
To address these two challenges, we adopt a moment generating function (MGF)-based stochastic network calculus approach \cite{fidler2010survey} for the analysis of networks of tandem queues. Then the service process, which is a function of the instantaneous channel capacity, is given in terms of the MGF of the fading channel distribution. This addresses the first challenge. Furthermore, we utilize network calculus to address the second challenge by using the service concatenation property. 

\subsection{Methodologies for Wireless Network Analysis}
Network calculus is an effective methodology for network performance analysis. It was originally proposed by Cruz \cite{cruz1991calculus, 61110} in the early 90's for the worst-case analysis of deterministic networked systems. 
Since then, the methodology has been extended to probabilistic settings.
Following the pioneering works in \cite{chang2000performance, chang2001performance, li2007network} on MGF-based traffic and service characterization and in order to model traffic and service processes with independent increments and to utilize  independence among multiplexed flows,  the  moment generating function (MGF) based  network calculus  was proposed \cite{fidler2010survey}.
Typically, the MGF approach to network calculus employs a finite-state Markov channel abstraction for the analysis of wireless fading channels \cite{Fidler-MGFQoS:2006, MGF-MIMO-Jiang:2011}. It is worth noting that  MGF-based approach was  used, outside the network calculus framework,  for the analysis of various fading channels and relaying channels, e.g., \cite{rao2014mgf, tabassum2014interference, alouini2000mgf, yilmaz2012unified}. In contrast, the $ (\min,\times) $ network calculus approach, proposed by \cite{alnetwork}, provides probabilistic performance bounds directly in terms of the fading channel parameters. It does that by transferring the problem from the `bit domain,' where traffic and service quantities are measured in bits, to the `SNR domain,' where these quantities are described by their SNR equivalence when measured at the channel capacity limit, using the exponential function.
To apply the $(\min, \times)$ network calculus to non-identically distributed multi-hop wireless networks, a recursive formula for delay bound computation was developed in \cite{petreskarecursive}. 


To evaluate the performance of our proposed outdoor $ 60 $ GHz wireless network, we model a multi-hop path in the network by a tandem of queues with service processes that represent the time-varying service offered by the underlying 60~GHz channel. Then we follow an MGF-based network calculus approach to compute probabilistic end-to-end delay and backlog bounds for that network. A particular specialty of our $60 $ GHz model relates to considering full-duplex relays to enable simultaneous transmission and reception. Although this leads to self-interference, full-duplex relaying can enhance the network throughput through the use of interference cancellation techniques \cite{duarte2010full, jain2011practical}.
Without loss of generality, we use a self-interference coefficient, which is a discounting parameter for the service offered by the channel, to characterize the interference at each relaying transceiver.

\subsection{Motivations and Contributions}
It is clear that although network calculus has been around for some years, its application to wireless networks analysis is fairly recent. Furthermore, in the existing related work, the self-interference factor was not taken into account. 
To our best knowledge, the performance guarantees of outdoor $ 60 $~GHz multi-hop wireless networks considering self-interfered channels have not been addressed before in the literature. Coupled with the importance of outdoor $ 60 $ GHz networks for the next generation mobile communications, it motivates us to investigate the backlog and delay performance as well as the constrained sum power budget and QoS trade-off corresponding to the self-interfered channel. 

The specific contributions of this paper are two-fold: (i) contribution to the theory of network calculus that is represented by a simplified closed-form expression for the network service curve applicable to both homogeneous and heterogeneous wireless networks, and 
(ii) contribution to the application by providing a service process characterization for the outdoor 60~GHz fading channel with self-interference, in terms of the MGF of the fading distribution.
Additional contributions of this work include
\begin{itemize}
\item An optimal power allocation scheme, based on the proposed methodology, more precisely, for 60~GHz multi-hop networks with i.i.d. shadowing under end-to-end delay constraint. 
\item The insight that, under optimal power allocation, the end-to-end performance bounds exponentially degrade with the self-interference coefficient. This suggests that managing self-interference can be extremely rewarding.
\end{itemize}
Our work builds on own previous work \cite{2016arXiv160800120Y}, where a service characterization for a single-hop 60 GHz system without self-interference was presented. 

The remainder of the paper is organized as follows. In Sec.~\ref{Sec: II}, we provide the basics for MGF-based stochastic network calculus. We construct a model for the outdoor $ 60 $ GHz wireless multi-hop network and derive its probabilistic backlog and delay bounds in Sec.~\ref{Sec: III}. In Sec.~\ref{Sec: IV} we propose an optimal power allocation strategy that results in better performance bounds. An asymptotic performance analysis of the network with  self-interference is presented in Sec.~\ref{Sec: V}. Numerical results and simulations are presented in Sec.~\ref{Sec: VI}, where we discuss the validity and  the effectiveness of our analytical upper bounds and investigate the impacts of self-interference coefficient and relay density on network performance.  Conclusions are presented in Sec.~\ref{Sec: VII}.

\section{Preliminaries}
\label{Sec: II}
In this section we mainly provide a brief review of network calculus fundamental results and the MGF-based stochastic network calculus framework in particular.
More details and the proofs for the presented fundamental results can be found for example in~\cite{le2001network,chang2000performance}.

\subsection{Model and Notation}
\label{Sec:Basic_Model}
\begin{figure}
\centering
\includegraphics[width=2.5in]{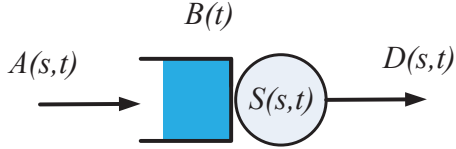}
\caption{A queuing model for a store-and-forward node.}
\label{fig: queuing model}
\end{figure}

Assuming a fluid-flow, discrete-time queuing system with a buffer of infinite size, and given a time interval $[s,t)$, $0\le s \leq t $, we define the non-decreasing (in $t$) bivariate processes $A(s,t), D(s,t)$ and $S(s,t)$ as the cumulative arrival to, departure from and service offered by the system as illustrated in Fig.~\ref{fig: queuing model}. 
We further assume that $A,D$ and $S$ are stationary non-negative random processes with $A(t,t)= D(t,t)= S(t,t)=0$ for all $t\ge 0$.
The cumulative arrival and service  processes  are given in terms of their instantaneous values during the $ i^{\mathrm{th}} $ time slot, $ a_{i} $ and $ s_{i} $ respectively, as follows
\begin{equation}\label{eq:cumulative-arrival-service}
A(s,t)=\sum_{k=s}^{t-1}a_{i}\; \mbox{ and } \; S(s,t)=\sum_{k=s}^{t-1}s_{i} \, ,
\end{equation}
for all $0 \le s \le t$. 
We assume time slots that are   normalized to 1 time unit. 
We denote by $B(t)$ the backlog (the amount of buffered data) at time $t$.
Furthermore, $W(t)$ denotes the virtual delay of the system at time $t$. 

Network calculus is based on $(\min,+)$-algebra, for which in particular the convolution and deconvolution are important to obtain bounds on the system performance. More precisely, given the non-decreasing and strictly positive bivariate processes $ X(s,t) $ and $ Y(s,t) $, the $(\min,+)$ convolution   and deconvolution are respectively defined as
\begin{equation*}
(X \otimes Y)(s,t) \triangleq \inf_{s\leq \tau \leq t}\{X(s,\tau)+Y(\tau,t) \},
\label{eqn:IIa1}
\end{equation*}
\begin{equation*}
(X \oslash Y)(s,t) \triangleq \sup_{0\leq\tau\leq s}\{X(\tau,t)-Y(\tau,s) \}.
\label{eqn:IIa2}
\end{equation*}

\subsection{Network Calculus Basics}
\label{sec: II_A}
In network calculus, the queuing system in Fig.~\ref{fig: queuing model} is analyzed with the arrival process $A(s,t)$ as input and the departure process $D(s,t)$ as system output.
Input and output are related to each through the $(\min,+)$ convolution of the input with the service process $S(s,t)$.
In particular, we consider in the following time varying systems known as \textit{dynamic servers}, for which for all $ t\geq 0 $ the network element offers a time varying service $S$ that satisfies the following input-output inequality~\cite{chang2000performance} $ D(0,t)\geq (A\otimes S)(0,t) $,
which holds with strict equality when the system is linear~\cite{le2001network}. 
One typical example is a work-conserving link with a time-variant capacity, with the available service $ S(s,t) $ during interval $ [s,t) $. 
Based on this server model, network calculus allows the study of the network performance in terms of total backlog and end-to-end delay. 
Both measures play critically important roles in system evaluation. 
On one hand, buffer dimensioning is a major factor to consider when designing and implementing broadband networks. 
This is true due to the space restriction and cost of  storage in intermediate network devices, e.g., routers in high data rate networks. 
On the other hand, end-to-end delay is closely related to the quality of service (QoS) and user experience for many networked applications, e.g., voice and video services. 
For a given queuing system with cumulative arrival $ A(0,t) $ and departure $ D(0,t) $ and for $t\ge 0$, the backlog at time $t$, $ B(t) $ is defined as the amount of traffic remaining in the system by time $t$. Therefore,
\begin{equation}\label{eqn:backlog1}
B(t)\triangleq A(0,t)-D(0,t)\, .
\end{equation}
Likewise, the virtual delay $W(t)$ is defined as the time it takes the last bit received by time $t$ to depart the system under a first-come-first-serve (FCFS) scheduling regime. Hence,
\begin{equation}\label{eqn:delay1}
W(t)\triangleq \inf\{w\geq 0: A(0,t)\leq D(0,t+w) \}.
\end{equation}
Substituting  $ D(0,t)\geq (A\otimes S)(0,t) $ in the above expressions and after some manipulation and using definitions of $ (\min,+) $ convolution and deconvolution, we can obtain the following bounds on $ B(t) $ and  $ W(t) $  respectively.
\begin{equation}
B(t)\leq (A\oslash S)(t,t),
\label{eqn:backlog}
\end{equation}
\begin{equation}
W(t) \leq \inf\{w\geq 0: (A\oslash S)(t+w,t)\leq 0 \}.
\label{eqn:delay}
\end{equation}

A main attribute of network calculus is its ability to handle  concatenated systems, e.g., multi-hop store-and-forward networks. 
This is mainly achieved using the server concatenation theory, which states that a network service process can be computed as the $(\min,+)$ convolution of the individual nodes' service processes~\cite{le2001network}.   
More precisely, given $ n $ tandem servers, let $ S_{i}(s,t) $ be the service process of the $ i^{\mathrm{th}}$ server, $ i\in \{1,\dots,n\} $, then the network service process $ S_{\mathrm{net}}(s,t) $ is given by
\begin{equation}\label{eqn:tandem server}
S_{\mathrm{net}}(s,t)=\left(S_{1}\otimes S_{2}\otimes \cdots \otimes S_{n} \right)(s,t) \,.
\end{equation}

\subsection{MGF-based Probabilistic Bounds}
\label{sec: II_B}
Deterministic network calculus~\cite{le2001network} can provide worst-case upper bounds on the backlog and the delay if traffic envelopes (an upper bound on the arrival process) as well as a service curve (a lower bound on the service process) are considered.
However, when analyzing systems with random input and/or service (like wireless networks), due to a possibly non-trivial probability for the service increment or arrival increment to be zero, the worst-case analysis is no longer useful to describe the performance any more. 
In such cases, probabilistic performance bounds provide more useful and realistic description of the system performance than worst-case analysis. 
%
%
%
%
In the probabilistic setting (where the arrival process $A$ and/or the service process $S$ are stationary random processes), the backlog and delay bounds defined in \eqref{eqn:backlog1} and \eqref{eqn:delay1} respectively are reformulated in a stochastic sense as:
\begin{equation}\label{eqn:probabilistic bounds}
\mathbb{P}\left(B(t)>b^{\eps'}\right)\leq \eps' \; \mbox{ and } \;
\mathbb{P}\left(W(t)>w^{\eps''}\right)\leq \eps'',
\end{equation}
where $ b^{\eps'} $ and $ w^{\eps''} $  denote the target probabilistic backlog and delay associated with violation probabilities $ \eps' $ and $ \eps'' $ respectively.
These performance bounds can be obtained by the distributions of the processes, i.e., in terms of the arrival and service processes MGFs \cite{fidler2010survey} or their Mellin transforms \cite{alnetwork}. 
These approaches constitute what we refer to as stochastic network calculus and they are most suitable for the analysis of wireless networks. 

In general, the MGF-based bounds are obtained by applying Chernoff's bound, that is, given a random variable $X$, we have  $ \mathbb{P}\left(X \geq x\right)\leq e^{-\theta x} \mathbb{E}\left[e^{\theta X}\right]=e^{-\theta x}\mathbb{M}_{X}(\theta) $, whenever the expectation exists, where $ \mathbb{E}\left[Y \right] $ and $ \mathbb{M}_{Y}\left(\theta\right) $ denote the expectation  and the moment generating function (or the Laplace transform) of $ Y $, respectively, and  $ \theta $ is an arbitrary non-negative free parameter. Given the stochastic process $ X(s,t), t\ge s $, we define the MGF of $ X $ for any $ \theta \geq 0 $ as  $ \mathbb{M}_{X}(\theta,s,t)\triangleq\mathbb{E}\left[e^{\theta X(s,t)}\right] $ \cite{fidler2006end}.
In addition, we define  $ \overline{\mathbb{M}}_{X}(\theta,s,t)\triangleq \mathbb{M}_{X}(-\theta,s,t)= \mathbb{E}\left[e^{-\theta X(s,t)}\right] $.

A number of properties of MGF-based network calculus are summarized in \cite{fidler2006end}. 
In this work, we consider a queuing system comprised of a set of tandem queues. Using \eqref{eqn:tandem server},  the MGF of the end-to-end service process, $ \overline{\mathbb{M}}_{S_{\mathrm{net}}}(\theta,s,t) $, of $N$ tandem queues with service processes $S_i, i = 1, \ldots N$, is bounded by
\begin{equation}\label{eq:network-MGF-service}
\overline{\mathbb{M}}_{S_{\mathrm{net}}}(\theta,s,t)\le 
\sum_{s \le u_1\le \dots \le u_N \le t} \prod_{i=1}^N \overline{\mathbb{M}}_{S_{i}}(\theta,u_{i-1},u_i)\, ,
\end{equation}
 where $u_0 = s$ and $u_N = t$.
In addition,  we denote the MGF of the arrival process by $ \mathbb{M}_{A}\left(\theta,s,t\right) $. Then probabilistic backlog and delay bounds satisfying \eqref{eqn:probabilistic bounds} above, can be respectively expressed by \cite{fidler2006end, al2013min}
\begin{equation}
b^{\eps'}=\inf_{\theta >0}\left\lbrace \frac{1}{\theta}\left(\log \mathsf{M}(\theta,t,t)-\log \eps' \right) \right\rbrace,
\label{eqn: backlog bound}
\end{equation}
\begin{equation}
w^{\eps''}=\inf\left\lbrace w : \inf_{\theta>0} \left\lbrace \mathsf{M}(\theta,t+w,t) \right\rbrace \leq \eps'' \right\rbrace,
\label{eqn: delay bound}
\end{equation}
where $ \mathsf{M}(\theta,s,t) $ is given as  
\begin{equation} 
\mathsf{M}(\theta,s,t)\triangleq\sum_{u=0}^{\min(s,t)}\mathbb{M}_{A}(\theta,u,t)\cdot \overline{\mathbb{M}}_{S_{\mathrm{net}}}(\theta,u,s).
\label{eqn: generic MGF for backlog and delay}
\end{equation}
It is worth noting that $ \mathsf{M}(\theta,s,t) $ is only valid when the arrival and service processes are independent. 
Also, we can see that, $ \mathsf{M}(\theta,s,t) $ given by \eqref{eqn: generic MGF for backlog and delay} is the key to derive the probabilistic backlog and delay bounds.

\section{Performance Analysis of Multi-Hop 60 GHz Wireless Network}
\label{Sec: III}

\subsection{System Model}

\begin{figure}
\centering
\includegraphics[width=3in]{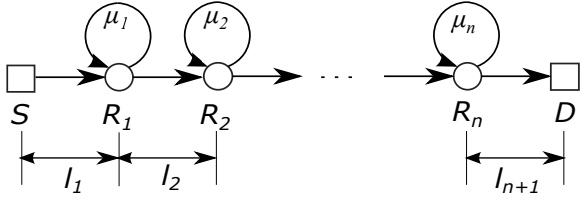}
\caption{A multi-hop $ 60 $~GHz wireless backhaul network with $ n $ relays.}
\label{fig: network model}
\end{figure}
We consider an outdoor $ 60 $ GHz wireless multi-hop relaying network  consisting of a source $ S $,  $ n $ ($ n \geq 1 $) relays $ R_{i} $, $ i=1,2,\ldots,n $ and a destination $ D $. 
For simplicity, we assign the labels $ 0 $, $ 1 $, $ \ldots $, $ n+1 $ to the ordered nodes. 
That is, $ S$ and $ D $  correspond to nodes $0$ and $ (n+1) $, respectively. 
Furthermore, we label the channel between nodes $ (i-1) $ and $ i $  as the $ i^{\mathrm{th}}$ hop in the set of hops $  \mathcal{I}_{\mathcal{H}}$, i.e., $ i \in \mathcal{I}_{\mathcal{H}}=\{1,2,\ldots,n+1\} $, and the distance between the two nodes by $ l_i \; \left[\mathrm{m}\right]$.
We denote the channel gain coefficient of the $ i^{\mathrm{th}} $ hop
by $ g_{i} $. Given the separation distance $ l_{i} $, a generalized model of $ g_{i} $ (in dB) for the 60 GHz outdoor channel is given by \cite{rangan2014millimeter, rappaport2015wideband}
\begin{equation}\label{eqn:channel gain}
g_{i}[\mathrm{dB}]=-\left(\alpha + 10\beta \log_{10} (l_i)+\xi_{i}\right),
\end{equation}
where $ \alpha $ and $ \beta $ are the least square fits of floating intercept and slope of the best fit, and $ \xi_{i}\sim \mathcal{N}(0,\sigma_{i}^2) $ corresponds to the log-normal shadowing effect with variance $ \sigma_{i}^2 $. The values of the parameters $ \alpha $ and $ \beta $ greatly depend on the  outdoor environment configurations. Our service characterization and performance analysis are carried out in terms of  these two parameters in order to incorporate all such configurations.

In addition to the (large scale) fading effects, the proposed model also considers self-interference at each full-duplex relay node.  The interference impact from other neighbouring nodes is small, due to the rapid attenuation of the millimeter waves, compared to self-interference and thus it is  ignored. 
A common approach to model the self-interference is to use a coefficient $ 0\leq \mu \leq 1 $ that characterizes the coupling between the transmitter and the receiver of a full-duplex device. It has been shown that the self-interference is linearly related the transmission power~\cite{duarte2010full}.

The signal to interference plus noise ratio (SINR) in the $ i^{\mathrm{th}} $ hop, denoted by $ \gamma_{i} $, for the described channel is expressed as 
\begin{align}
\gamma_{i}=&\kappa \cdot\omega_{i}\cdot g_{i} \, , \notag \\
&\, s.t., \, \,\,\, \omega_{i}=
\begin{dcases}
\frac{  \lambda_{i-1}}{1+\mu_{i} \lambda_{i} }, & i\in \{1,2,\ldots,n\}\\
\lambda_{i-1}, & i=n+1
\end{dcases},
\label{eqn: SINR}
\end{align}
 where, $ \kappa $ is a scalar depending on system configuration, i.e., the antenna gains of the communication pair, $ g_{i} $ is the channel gain coefficient given by \eqref{eqn:channel gain}, $\lambda_i \triangleq \frac{P_{i}}{N_{0}}  $ denotes the transmitted signal-to-noise ratio (SNR) at node $ i $ corresponding  to  transmit power $ P_{i} $ and background noise power $ N_{0} $.

For the multi-hop scenario, we assume the stochastic process of each hop to be stationary and independent in time. 
That is, we can use a series of independent random variables $ \gamma_i$ to characterize the multi-hop channels, namely, $ \gamma_{i}^{(k)} \overset{\ell}{=} \gamma_{i}$ in all time slot $ k $, where $\overset{\ell}{=}$ denotes equality in law (i.e., in distribution). 
The shadowing effect, which is due to objects obstructing the propagation of $ 60 $ GHz radios, is considered in the channel gain coefficient model given by~\eqref{eqn:channel gain}. 
Regarding the stochastic behavior of different links, generally, shadowing is not spatially independent. 
Since highly directional antennas for $ 60 $~GHz are assumed, it is safe to assume that the obstructions of radio propagation behave independently, which justifies the independence assumption across hops. 

In general, the fading distributions of the subsequent channels in multi-hop wireless network may not be identical. Nevertheless, it is worthwhile to decompose the set of hops into subsets of hops with identically distributed channel gains.  
More precisely, we decompose the set of hops, $ \mathcal{I}_{\mathcal{H}}  $, into $m$  subsets, $\mathcal{X}_{k}, k \in \mathcal{I}_{\mathcal{M}} =\{ 1, \dots, m\}$, where, $\mathcal{I}_{\mathcal{H}}=\bigcup_{k=1}^{m} \mathcal{X}_{k}$, with $ \mathcal{X}_{i}\bigcap \mathcal{X}_{j}=\emptyset $ for all $ i,j \in \mathcal{I}_{\mathcal{M}} $ such that $ i\neq j $,  where $ \mathcal{X}_{k} $ is defined as the set of indices such that
\begin{equation}\label{eqn: set of X}
\mathcal{X}_{k}=\{j\in \mathcal{I}_{\mathcal{H}}, k \in \mathcal{I}_{\mathcal{M}}: F_{\gamma_j}(x) {=} F^{\langle k \rangle}(x)\},
\end{equation}
where $F_{X}(x)$ is the probability distribution function of the random variable $X$,  $ F^{\langle k \rangle}(x) $ represents a unique distribution function corresponding to the subset of i.i.d. hops denoted by the index $k \in \mathcal{I}_{\mathcal{M}} $. We emphasize that $ |\mathcal{I}_{\mathcal{H}} | \ge |\mathcal{I}_{\mathcal{M}}|   $, where $|\mathcal{Y}|$ represents the  cardinality of the set $\mathcal{Y}$, and the equality is attained when the multi-hop network is fully heterogeneous.  In such extreme case, each subset $ \mathcal{X}_{k} $ contains only one element, i.e., the channel at each hop has  different fading distribution than that at the other hops. 

To address transmit power allocation for the multi-hop network, we consider a system with  sum power $P_{\mathrm{tot}}$ constraint, i.e., $ \sum_{i=0}^{n}P_{i} = P_{\mathrm{tot}}$. 
Equivalently, given a constant background noise power $ N_{0} $ for all hops,  the sum power constraint can be reformulated as
\begin{equation}\label{eq: power constraint}
\sum_{i=0}^{n} \lambda_{i}= \lambda_{\mathrm{tot}}\triangleq \frac{P_{\mathrm{tot}}}{N_{0}}.
\end{equation}

\subsection{MGF Bound for the Cumulative Service Process}
The performance bounds given by \eqref{eqn: backlog bound} and \eqref{eqn: delay bound} require the computation of MGFs of arrival and service processes. 
For the arrival process we consider in this work the $ \left(\sigma(\theta),\rho(\theta)\right) $ traffic characterization with parameters $ \sigma(\theta)=\delta_b $ and $ \rho(\theta)=\rho_a $, i.e., we assume deterministically bounded arrival process. 
This gives an upper bound of the MGF of the arrival process as
\begin{equation} 
\mathbb{M}_{A}(\theta,s,t)\leq 
e^{\theta \delta_b}\left(p_{a}(\theta)\right)^{t-s},
\label{eqn: MGF Arrival Process}
\end{equation}
for any $\theta > 0 $, where $ p_{a}(\theta)=e^{\theta\rho_{a}} $.

Regarding the service process, we present in the following a series of results that apply to a Shannon-capacity type service process.
Given a channel SINR $ \gamma $, in this model the service process of the channel is given by 
$C(\gamma)=\eta\ln (1+\gamma)$
 bits/s, where $ \eta=\frac{W}{\ln 2} $ with channel bandwidth $ W $. 
By \eqref{eq:cumulative-arrival-service}, 
the  cumulative service process  for hop $i$, $ S_{i}(s,t) $, is given by 
\begin{equation}\label{eq:MGF-service}
S_{i}(s,t) = \sum_{k=s}^{t-1} C(\gamma_i(k)) = \eta\sum_{k=s}^{t-1}\ln \left(1+\gamma_{i}(k)\right),
\end{equation}
where $ \gamma_{i}(k) $ is the instantaneous SINR in the $ k^{\mathrm{th}} $ time slot for the $ i^{\mathrm{th}} $ hop given in terms of $g_i$ in \eqref{eqn:channel gain} and $\omega_i$ in \eqref{eqn: SINR}.

An exact expression for the MGF of this Shannon-type cumulative service process in~\eqref{eq:MGF-service} is intractable.
Instead, in the following lemma we present bounds on the MGF of such Shannon-type service processes $ S_{i}(s,t) $:
\begin{lemma}\label{lemma:inverse moment}
Let $ F_{X}(x) $ denote the cumulative distribution function (c.d.f.) of a non-negative random variable $ X $, for any $ \delta >0 $ and $ \theta \geq 0 $ we have $$   \mathbb{E}\left[(1+X)^{-\theta}\right]
\leq \mathcal{U}_{\delta,X}(\theta)$$,
where
\begin{equation*}
\mathcal{U}_{\delta,X}(\theta)
= \min_{u\geq 0} \left\lbrace \left(1+\delta N_{\delta}(u) \right)^{-\theta} + \sum_{k=1}^{ N_{\delta}(u)} a_{\theta,\delta}(k) F_{X}(k\delta) \right\rbrace \,,
\end{equation*}
where $ N_{\delta}(u) $ and $ a_{\theta,\delta}(k) $ are respectively given by $ N_{\delta}(u)=\lfloor \frac{u}{\delta}\rfloor $ and $ a_{\theta,\delta}(k)= \left(1+(k-1)\delta\right)^{-\theta}-\left(1+k\delta\right)^{-\theta}$.
\end{lemma}
For the proof of Lemma~\ref{lemma:inverse moment}, please refer to \cite{2016arXiv160800120Y}.
Note that the tightness of the bound obtained in Lemma \ref{lemma:inverse moment} depends on the parameter $ \delta $, the discretization step size. Technically, a smaller step size yields a tighter upper bound while leading to higher computational costs.

Based on Lemma \ref{lemma:inverse moment}, a bound on the MGF of the service process for any single-server wireless system with service process increments given by the Shannon capacity is given:
\begin{theorem}\label{theorem:upper bound of M_S}
Given the service process $ S(s,t)=\eta\sum_{k=s}^{t-1}\ln \left(1+\gamma(k)\right) $ with independent positive $ \gamma(k) $, an upper bound on $ \overline{\mathbb{M}}_{S}(\theta,s,t) $ is given by
\begin{equation*}
\overline{\mathbb{M}}_{S}(\theta,s,t) \leq \prod_{k=s}^{t-1}q(\theta,k), \\
\end{equation*}
where $ q(\theta,k) = \mathcal{U}_{\delta,\gamma(k)}\left(\eta\theta\right)$. Furthermore, if for all $ k $,  $ \gamma(k) \overset{\ell}{=} \gamma $   then $ q(\theta,k)=q(\theta) $   and the above expression  reduces to $ \overline{\mathbb{M}}_{S}(\theta,s,t) \leq \left ( q(\theta) \right )^{t-s}$.
\end{theorem}

\begin{IEEEproof}
Starting from the definition of $ \overline{\mathbb{M}}_{S}(\theta,s,t) $ and using the independence assumption of $ \gamma(k) $ in $k$, we have
\begin{equation*}
\begin{split}
& \overline{\mathbb{M}}_{S}(\theta,s,t)= \mathbb{E}\left[\exp\left(-\theta \cdot S(s,t)\right)\right]\\
= & \mathbb{E}\left[\prod_{k=s}^{t-1}\exp\left(-\theta \eta \ln (1+\gamma(k))\right)\right]
= \prod_{k=s}^{t-1}\mathbb{E}\left[\left(1+\gamma(k)\right)^{-\theta \eta}\right].
\end{split}
\end{equation*}

Applying Lemma \ref{lemma:inverse moment} to the right hand side of the expression above, Theorem \ref{theorem:upper bound of M_S} immediately follows.
\end{IEEEproof}

Next, we provide a MGF bound on the network service for multi-hop wireless networks with heterogeneous, independent Shannon-type service processes per link.  The channel categorization, shown in \eqref{eqn: set of X}, is used for expression simplifications. 

\begin{theorem}\label{theorem:upper bound of M_S_net}
The network service process $ S_{\mathrm{net}}(s,t) $ of a multi-hop wireless network consisting of $ n $ relays and characterized by the decomposable set of hops $\mathcal{I}_{\mathcal{H}}=\bigcup_{i=1}^{m} \mathcal{X}_{i}$ following \eqref{eqn: set of X}, where the subset  $ \mathcal{X}_i $ is associated with the randomly varying SINR $ \hat{\gamma}_i $ and has a Shannon-type service process increment $\ln \left(1+\gamma\right) $, has the following MGF bound
\begin{align}\label{eqn: MGF Service_Net-m1}
\overline{\mathbb{M}}_{S_{\mathrm{net}}}(\theta,s,t) \le
\sum_{\sum\limits_{i=1}^{m}\pi_i=t-s} \prod_{i=1}^{m}\binom{\pi_{i}+|\mathcal{X}_{i}|-1}{|\mathcal{X}_{i}|-1}\hat{q}_{i}^{\pi_{i}}(\theta),
\end{align}
where  $ \hat{q}_{i}(\theta)=\mathcal{U}_{\delta,\hat{\gamma}_i}(\eta\theta) $ for all $ i\in \mathcal{I}_{\mathcal{M}}  $.
\end{theorem}
\begin{IEEEproof}
Using equation \eqref{eqn:tandem server}, we can bound the MGF  for the $n+1$ hops network service process by 
\begin{align}
 \overline{\mathbb{M}}_{S_{\mathrm{net}}}(\theta,s,t)
{=}& \left( \overline{\mathbb{M}}_{S_{1} \otimes  \dots \otimes S_{n+1}} \right)(\theta,s,t) \notag \\
{\leq} & \sum_{s=\tau_1\le \dots \le \tau_{n+1} = t} \,\, \prod_{i=1}^{n+1} \overline{\mathbb{M}}_{S_{i}}(\theta,\tau_{i-1},\tau_i)  \notag \\
{\leq}& 
\sum_{\sum_{i=1}^{n+1}\tau_i =t-s} q_{1}^{\tau_1}(\theta)q_{2}^{\tau_2}(\theta)\cdots q_{n+1}^{\tau_{n+1}}(\theta) \notag \\
{=} & \sum_{\sum_{i=1}^{m}\pi_i=t-s} \,\, \prod_{i=1}^{m}\hat{q}_{i}^{\pi_i}(\theta)   \sum_{\sum_{k \in \mathcal{X}_i} \tau_k = \pi_i} 1    \notag 
\label{eq:Th2-11}
\end{align}
where  the first inequality is obtained by using \eqref{eq:network-MGF-service} and the second inequality   is obtained by using the change of variables $\tau_i = \tau_i - \tau_{i-1}$ and the stationarity  of the service processes, i.e.,  $\overline{\mathbb{M}}_{S_{i}}(\theta,\tau_{i-1},\tau_i) = \overline{\mathbb{M}}_{S_{i}}(\theta,\tau_i-\tau_{i-1}) $, then  applying    Theorem~\ref{theorem:upper bound of M_S}. The equality in the last line is obtained by aggregating similar terms, i.e., $\pi_i = \sum_{k \in \mathcal{X}_i} \tau_k$. Applying the combinations of multisets theory \cite{brualdi1992introductory}, it is known that
\begin{equation*}
\sum_{\sum_{k \in \mathcal{X}_i} \tau_k = \pi_i} 1 = \binom{\pi_{i}+|\mathcal{X}_{i}|-1}{|\mathcal{X}_{i}|-1},
\end{equation*}
where $ |\mathcal{X}_i| $ denotes the cardinality of $ \mathcal{X}_i $, and then the theorem is concluded.
\end{IEEEproof}

We emphasize that all results presented so far apply to general cases of distributions of link SINR $\gamma$.
Thus, the results have wide applicability to wireless (and wired) network analysis, as long as Shannon-type service processes are assumed.

\subsection{Probabilistic Performance Bounds} 
The general probabilistic total backlog and end-to-end delay bounds for a multi-hop wireless network are given by \eqref{eqn: backlog bound} and \eqref{eqn: delay bound}, respectively.  Both bounds are given in terms of $ \mathsf{M}(\theta,s,t) $ in \eqref{eqn: generic MGF for backlog and delay}.
Theorem~\ref{theorem:upper bound of M} provides an upper bound on the function $ \mathsf{M}(\theta,s,t) $, and hence, probabilistic performance bounds for multi-hop wireless networks, when the arrival is characterized by \eqref{eqn: MGF Arrival Process} and the network service is provided by Theorem \ref{theorem:upper bound of M_S_net}. 

Let us first define $ \mathcal{K}_{\tau,n,m}\left(x\right) $ as
\begin{align}
&\mathcal{K}_{\tau,n,m}\left(x\right) \notag \\
 \triangleq & x^\tau \binom{n+1-m+\tau}{n+1-m}
 \cdot {}_2 F_{1}\left(1,n+2-m+\tau; \tau+1;x\right), \notag
\end{align}
where the \emph{Generalized Hypergeometric Function} $ {}_p F_{q}\left(\underline{a};\underline{b};x\right) $, with vectors $ \underline{a}=\left[a_1,\ldots,a_p\right] $ and $ \underline{b}=\left[b_1,\ldots,b_q\right] $, is given as
\begin{equation*}
{}_p F_{q}\left(\underline{a}; \underline{b};x\right) \triangleq \sum_{k=0}^{\infty}
\left(\dfrac{\prod_{i=1}^{p}\left(a_i\right)_k}{\prod_{j=1}^{q}\left(b_j\right)_k}\right) \cdot\frac{x^k}{k!},
\end{equation*}
and $ (a_i) _k$ and $ (b_i)_k $ are \emph{Pochhammer symbols}.

%
\begin{theorem}\label{theorem:upper bound of M} 
Let $ m $ be the number of subsets of identically distributed channels, $ \tau\triangleq\max(s-t,0)$, and  $ V_i\left(\theta\right) \triangleq  p_a(\theta)\hat{q}_{i}(\theta)$, a upper bound for  $ \mathsf{M}(\theta,s,t) $,  $\theta > 0$, for the $ \left(n+1\right) $-hop wireless network is given by 
\begin{equation*}
\mathsf{M}(\theta,s,t) \leq \frac{e^{\theta \delta_b}}{p_{a}^{s\!-\!t}(\theta)}\! \sum_{i=1}^{m}\psi_{i}(\theta) V_{i}^{m-1}(\theta) \mathcal{K}_{\tau,n,m}\left(V_i\left(\theta\right)\right),
\end{equation*}
whenever the stability condition, 
$ \max\limits_{i \in \mathcal I_{\mathcal M}} \lbrace V_i(\theta)\rbrace <1 $, is satisfied.
Here, $ \psi_i\left(\theta\right) $ for all $ i \in \{1,\ldots,m\} $ is defined as
\begin{equation}\label{eq:Thm3-psi}
\psi_i\left(\theta\right) \triangleq 
\begin{dcases}
\prod_{j \neq i} {\left(V_{i}(\theta)-V_{j}(\theta)\right)^{-1}}, & m \geq 2 \\
1,& m=1
\end{dcases}.
\end{equation}
\end{theorem}
\begin{IEEEproof} 
By the definition of $ \mathsf{M}\left(\theta,s,t\right) $ in \eqref{eqn: generic MGF for backlog and delay}, we start with the  substitution of \eqref{eqn: MGF Arrival Process} and \eqref{eqn: MGF Service_Net-m1} in \eqref{eqn: generic MGF for backlog and delay},  then we have
\begin{align}
& \mathsf{M}(\theta,s,t)  \notag \\
\leq & e^{\theta \delta_b}   \sum_{u=0}^{\min(s,t)}   p_{a}^{t-u}(\theta) \cdot \sum_{\sum\limits_{i=1}^{m}\pi_i=s-u} \prod_{i=1}^{m}\binom{\pi_{i}+|\mathcal{X}_{i}|-1}{|\mathcal{X}_{i}|-1}\hat{q}_{i}^{\pi_{i}}(\theta). \notag
\end{align}
Then using the change of variable $ u=s-u $ and rearranging terms, we equivalently have
\begin{align}
& \mathsf{M}(\theta,s,t) \notag \\
\leq & \frac{e^{\theta \delta_b}}{p_{a}^{s-t}(\theta)} \sum_{u=\tau}^{s}   p_{a}^{u}(\theta) 
\cdot \sum_{\sum\limits_{i=1}^{m}\pi_i=u} \prod_{i=1}^{m}\binom{\pi_{i}+|\mathcal{X}_{i}|-1}{|\mathcal{X}_{i}|-1}\hat{q}_{i}^{\pi_{i}}(\theta), \notag
\end{align}

Based on above, by splitting the power $ u $ for $ p_a^u\left(\theta\right) $ into components in terms of $ \pi_i $, we immediately have
\begin{align}
\mathsf{M}(\theta,s,t) 
\leq & \frac{e^{\theta \delta_b}}{p_{a}^{s-t}(\theta)} \sum_{u=\tau}^{s} \sum_{\sum\limits_{i=1}^{m}\pi_i=u} \prod_{i=1}^{m}\binom{\pi_{i} \!+\!|\mathcal{X}_{i}| \!-\! 1}{|\mathcal{X}_{i}| \!-\! 1}V_i^{\pi_{i}}\left(\theta\right) \notag \\
\leq & \frac{e^{\theta \delta_b}}{p_{a}^{s-t}(\theta)} \sum_{u=\tau}^{\infty} \sum_{\sum\limits_{i=1}^{m}\pi_i=u} \prod_{i=1}^{m}\binom{\pi_{i} \!+\!|\mathcal{X}_{i}| \!-\! 1}{|\mathcal{X}_{i}| \!-\! 1}V_i^{\pi_{i}}\left(\theta\right), \notag
\end{align}
where $ V_i\left(\theta\right) \triangleq  p_a(\theta)\hat{q}_{i}(\theta)$ and $ \tau \triangleq \max\left(s-t,0\right) $ are respectively defined for notional simplicity, and the last inequality is obtained by pushing $ s $ to infinity. Then, we further have
\begin{align}\label{eq:Thm3-1}
& \mathsf{M}\left(\theta,s,t\right) \notag \\
\leq & \frac{e^{\theta \delta_b}}{p_{a}^{s-t}(\theta)} \sum_{u=\tau}^{\infty} \binom{u+n+1-m}{n+1-m}  \sum_{\sum\limits_{i=1}^{m}\pi_i=u} \prod_{i=1}^{m}V_i^{\pi_{i}}\left(\theta\right),
\end{align}
where the following inequality for combinatorics is used, i.e.,
\begin{equation*}
\binom{n_1}{k_1}\binom{n_2}{k_2}\cdots\binom{n_M}{k_M}\leq \binom{n_1+n_2+\cdots+n_M}{k_1+k_2+\cdots+k_M}
\end{equation*}
for all $ n_i\geq k_i\geq 0 $, $ i\in \{1,2,\ldots,M\} $. 

From \eqref{eq:Thm3-1} on, we need to consider two situations: (i) $ m=1 $ for the homogeneous network, and (ii) $ m \geq 2 $, which represents the heterogeneous scenario. 

For $ m=1 $, the expression in \eqref{eq:Thm3-1} directly reduces to
\begin{align}\label{eq:Thm3-2}
\mathsf{M}\left(\theta,s,t\right) \leq \frac{e^{\theta \delta_b}}{p_{a}^{s-t}(\theta)} \sum_{u=\tau}^{\infty} \binom{u+n}{n}  V^{u}\left(\theta\right),
\end{align}
where $ V\left(\theta\right) = p_a\left(\theta\right)\hat{q}\left(\theta\right) $, and $ \hat{q}\left(\theta\right) $ characterizes the homogeneous channel gain.

Regarding $ m\geq 2 $, the upper bound of $ \mathsf{M}\left(\theta,s,t\right) $ can be formulated as
\begin{align}\label{eq:Thm3-3}
& \mathsf{M}\left(\theta,s,t\right) \notag \\
\leq & \frac{e^{\theta \delta_b}}{p_{a}^{s-t}(\theta)} \sum_{u=\tau}^{\infty} \binom{u\! + \! n \! + \! 1 \! - \! m}{n \! + \! 1 \! - \! m}  \sum_{i=1}^{m}\varphi_i\left(\theta\right) V_i^{u+m-1}\left(\theta\right)  \\
= & \frac{e^{\theta \delta_b}}{p_{a}^{s-t}(\theta)} \sum_{i=1}^{m} \varphi_i\left(\theta\right) V_i^{m-1}\left(\theta\right) \sum_{u=\tau}^{\infty} \binom{u\! + \! n \! + \! 1 \! - \! m}{n \! + \! 1 \! - \! m} V_i^{u}\left(\theta\right) \notag,
\end{align}
where $\varphi_i\left(\theta\right) = \prod_{j \neq i} {\left(V_{i}(\theta)-V_{j}(\theta)\right)^{-1}} $. Here, the inequality comes from the application of following homogeneous polynomials identity \cite{belbachir2006linear}, that is, 
\begin{equation*}
\sum_{k_1+\cdots+k_M=N}x_{1}^{k_{1}}x_{2}^{k_{2}}\cdots x_{M}^{k_{M}} = \sum_{i=1}^{M} \dfrac{x_{i}^{N+M-1}}{\prod_{j\neq i}(x_{i}-x_{j})}
\end{equation*}
for any $ M \ge 2 $ distinct variables, $ x_1,x_2,\ldots,x_M $.

Furthermore, we have a closed-form expression in terms of generalized hypergeometric function for the infinite sum \cite{wolfram2010wolframalpha}
\begin{equation*}
\sum_{k=u}^{\infty}\binom{n+k}{k}x^{k}=x^{u}\binom{n+u}{u} {}_{2}F_{1}\left(1,n+u+1;u+1;x\right),
\end{equation*}
whenever $ |x|<1 $, which subsequently produces the newly defined $ \mathcal{K}_{\tau,n,m}\left(x\right) $. It is worth noting that, the condition $ \max\limits_{i \in \mathcal I_{\mathcal M}} \lbrace V_i(\theta) \rbrace < 1$ should be satisfied, for the sake of stability.
Combining \eqref{eq:Thm3-2} and \eqref{eq:Thm3-3} and using \eqref{eq:Thm3-psi}, then the theorem can be concluded.
\end{IEEEproof}

The stability condition in Theorem \ref{theorem:upper bound of M} can be reasoned as follows. If we take the $\log$ of both sides of the expression, the condition can be stated as, ``the difference between the $\log$~MGF of the arrival and that of the service is less than 0,'' i.e., for a given QoS measure $\theta$, the effective capacity must exceed the effective bandwidth, for the same $\theta$, for the system to be stable. This intuitive result was hinted in \cite[Ch.~7]{chang2000performance}.  

It is evident that, the upper bound of $ \mathsf{M}\left(\theta,s,t\right) $ by Theorem~\ref{theorem:upper bound of M}  is suitable for both  homogeneous and heterogeneous multi-hop wireless networks. 
In contrast to the recursive method \cite{petreskarecursive},  the classification of hops of the network into $m$ subsets with identical channels, in terms of their channel distributions, provides a straightforward approach to compute $ \mathsf{M}\left(\theta,s,t\right) $, thereby largely simplifying the obtained expression, 
since from the side of service process each subset of nodes can be considered as homogeneous sub-network, 
and the overall network service process is given by the $(\min, +)$~convolution of the subnets service processes. 


Clearly, the expression provided by Theorem~\ref{theorem:upper bound of M} depends on the generalized hypergeometric function, which is computational costly. 
In what follows, we  provide a looser but more simplified upper bound on $ \mathsf{M}\left(\theta,s,t\right)  $ in the homogeneous case specifically, i.e., $m=1$. 
For that, we first need the following lemma regarding the upper bound on $ \mathcal{K}_{\tau,n,1}\left(x\right) $:

\begin{lemma}\label{lemma:combinatorics sum}
For non-negative integers $ n $ and $  \tau $, the inequality 
$$ \mathcal{K}_{\tau,n,1}\left(x\right) \le  \mathcal{G}_{\tau,n}(x)\triangleq  \min\left\lbrace\mathcal{G}_1(x),\mathcal{G}_2(x)\right\rbrace $$ holds for $ 0\leq x < 1 $, where $ \mathcal{G}_1(x) $ and $ \mathcal{G}_2(x) $ are respectively given by
\begin{equation*}
\mathcal{G}_{1}(x)=\frac{\min \left(1,x^\tau  \binom{n+\tau}{n}  \right)}{(1-x)^{n+1}}
\end{equation*}
and
\begin{equation*}
\mathcal{G}_2(x)=\frac{1}{(1-x)^{n+1}}-\binom{n+\tau}{n+ 1}x^{\tau-1}.
\end{equation*}
\end{lemma}
\begin{IEEEproof}
Note that $ \mathcal{K}_{\tau,n,1}\left(x\right) $ is explicitly formulated as
$$\mathcal{K}_{\tau,n,1}\left(x\right) = \sum_{k=\tau}^{\infty}\binom{n+k}{k}x^{k}, $$
then the derivation can be performed from the the following two cases:

(i) It is easy to know that, $ i+k\leq(i+k-\tau)(1+\frac{\tau}{i}) $ holds for all $ i\geq 1 $ and all $ k\geq \tau $, then we have
\begin{equation*}
\binom{n+k}{k}\leq \binom{n+k-\tau}{k-\tau}\binom{n+\tau}{n}.
\end{equation*}
Then we have
\begin{equation}\label{eq:Lem1-2}
\mathcal{K}_{\tau,n,1}\left(x\right) \leq x^\tau \binom{n+\tau}{n} \sum_{k=0}^{\infty}\binom{n+k}{k}x^k,
\end{equation}
where we used the change of variables, i.e., $k = k - \tau$. It immediately gives $ \mathcal{G}_1\left(x\right) $ by applying \emph{Newton's Generalized Binomial Theorem} and by taking $ 1 $  as the upper limit into account as well. 

(ii) On the other hand, it is easy to know that
\begin{equation*}
\begin{split}
\sum_{k=\tau}^{\infty}\binom{n+k}{k}x^{k}
= & \frac{1}{(1-x)^{n+1}}-\sum_{k=0}^{\tau-1}\binom{n+k}{k}x^{k}\\
\leq & \frac{1}{(1-x)^{n+1}}- x^{\tau-1}\sum_{k=0}^{\tau-1}\binom{n+k}{k}\\
= & \frac{1}{(1-x)^{n+1}}- \binom{n+\tau}{n+1}x^{\tau-1},
\end{split}
\end{equation*}
where inequality is true since $ 0\leq x<1 $, and the last equality is obtained by applying a combinatorial property \cite{brualdi1992introductory}.  Then $ \mathcal{G}_{2}\left(x\right) $ is obtained.
\end{IEEEproof}

The lemma above allows us to use a simpler expression to describe the upper bound (compared to Theorem~\ref{theorem:upper bound of M}) on $ \mathsf{M}\left(\theta,s,t\right) $ for homogeneous networks in particular, and it is shown in the following theorem.  
\begin{theorem}\label{theorem:homogeneous M}
For homogeneous $ \left(n+1\right) $-hop wireless networks characterized by the MGF service bound $\hat{q}(\theta) $, and for any $ \theta > 0 $,  given $ p_a\left(\theta\right) $ we have
\begin{equation*}
\mathsf{M}(\theta,s,t) \leq \frac{e^{\theta \delta_b}}{p_a^{s-t}(\theta)} 
\cdot
\mathcal{G}_{\tau,n}\left(p_a(\theta)\hat{q}(\theta)\right), 
\end{equation*}
where $ \tau\triangleq\max(s-t,0)$, whenever the stability condition, 
$p_a(\theta)\hat{q}(\theta)<1$, holds.
\end{theorem}
\begin{IEEEproof}
Applying Lemma \ref{lemma:combinatorics sum} in \eqref{eq:Thm3-2} whenever $ 0 \le p_a(\theta)\hat{q}(\theta)<1 $, then Theorem~\ref{theorem:homogeneous M} follows.
\end{IEEEproof}

Inserting the results from Theorem~\ref{theorem:upper bound of M} (or Theorem~\ref{theorem:homogeneous M} for merely homogeneous cases),  in \eqref{eqn: backlog bound} and \eqref{eqn: delay bound}, we obtain the desired probabilistic bounds on backlog $ b^{\eps} $ and delay $ w^{\eps} $ in the network, respectively. 

\section{Power Allocation for 60 GHz Networks}
\label{Sec: IV}

In this section, we study optimal power allocation for a multi-hop 60 GHz multi-hop network, applying the results presented in the previous section regarding the MGF-based network calculus. 
More precisely, we are interested in finding the optimal transmit power allocation of multi-hop network with independent and identically distributed shadowing per hop,  $ \xi_{i}, \, \forall i \in \mathcal{I_H} $, assuming that $ \xi_{i}$ is stationary. 
In particular, we limit our study to the case of homogenous log-normally distributed shadowing over all links, i.e. we set $ \xi_{i}\overset{\ell}{=}\xi\sim \mathcal{N}(0,\sigma^2) $ for all hops $i \in \mathcal{I_H}$. 
The case with non-identical shadowing is more involved and is left for future work. 

Equations \eqref{eqn: backlog bound} and \eqref{eqn: delay bound} show that  probabilistic bounds on the total network backlog and end-to-end delay performance are determined in terms of the function $ \mathsf{M}(\theta,s,t) $ defined in \eqref{eqn: generic MGF for backlog and delay}. This implies that the performance optimization, e.g., with respect to transmit power allocation, of a multi-hop network  is equivalent to optimizing the function $ \mathsf{M}(\theta,s,t) $, which, with a given arrival process $ \mathbb{M}_{A} $, is equivalent to optimizing $ \overline{\mathbb{M}}_{S_{\mathrm{net}}} $. Therefore, in what follows, the optimization subject is  the MGF  for the network service process denoted by $ \overline{\mathbb{M}}_{S_{\mathrm{net}}} $.  Since an exact expression for the MGF of network service process is not attainable, we opt for optimizing a bound on $ \overline{\mathbb{M}}_{S_{\mathrm{net}}} $, given by \eqref{eq:network-MGF-service}, instead. A power allocation that elevates the lower bound on network service capacity corresponds to lower $ \overline{\mathbb{M}}_{S_{\mathrm{net}}} $ and thus reduces the probabilistic bounds on network performance.

Theorem~\ref{theorem:upper bound of M_S_net} provides an upper bound on $ \overline{\mathbb{M}}_{S_{\mathrm{net}}}(\theta,s,t) $ and hence a lower bound on the network's service capability. Therefore, in order to maximize the lower bound on the service process, we must minimize the upper bound on $ \overline{\mathbb{M}}_{S_{\mathrm{net}}}(\theta,s,t) $. 
The function $ \overline{\mathbb{M}}_{S_{\mathrm{net}}}(\theta,s,t) $ is related to the power allocation vector $\mathbf{P} \in \mathbb{R}_{+}^{n+1}  $. From \eqref{eq:network-MGF-service} we have 
\begin{equation}\label{eqn:lower bound of service capability}
\overline{\mathbb{M}}_{S_{\mathrm{net}}}(\theta,s,t)
\leq  \sum_{\sum_{i=1}^{n+1}\pi_{i}=t-s}\prod_{i=1}^{n+1}\left(\mathbb{E}\left[(1+\gamma_{i})^{-\theta \eta}\right]\right)^{\pi_{i}},
\end{equation}
where the per node $i$ SINR, $ \gamma_{i} $, $ \forall i\in \mathcal{I}_{\mathcal{H}} $, is given by \eqref{eq:MGF-service} and is in turn related to the allocated transmit power for the corresponding node.
Let $ \mathbf{\Xi}_{n} \subset \mathbb{R}_{+}^{n+1}$ be the set of feasible power allocation schemes with respect to $ n $ intermediate nodes. Furthermore, the sum of all allocated power should be constrained by the total power budget, i.e., $$ P_{\sum}\triangleq\sum_{ i\in \mathcal{I}_{\mathcal{H}}}  P_i \le  P_{\rm tot}, $$
where $ P_{\rm tot}$ is the total power budget. For any power allocation $ \mathbf{P}\triangleq \left\lbrace P_{i}\right\rbrace_{i\in \mathcal{I}_{\mathcal{H}} } \in \mathbf{\Xi}_{n} $, we define $ \mathbf{P} $ as a \emph{feasible} power allocation scheme if the power constraint above is satisfied.

To determine the optimal power allocation strategy, we need the following two lemmas first. To be more precise, in  Lemma~\ref{lemma:3} below, we show that given a feasible power allocation, the maximal network service capability is achieved only when the SINRs for all hops are identically distributed.
Next, in Lemma~\ref{lemma:4}  we show the existence and uniqueness of the optimal power allocation vector $ \mathbf{P}^* $ when $ P_{\sum} = P_{\rm tot} $.

\begin{lemma}[Sufficiency]\label{lemma:3}
Given the sum transmit power budget $ P_{\rm tot} $ for a $\left(n+1\right)$-hop wireless network,  a feasible power allocation $ \mathbf{P}^* $, where $ P_{\sum}=P_{\rm tot} $, that results in identically distributed SINR over all hops maximizes the lower bound on network service process whenever such $ \mathbf{P}^* $ exists.
\end{lemma}
\begin{IEEEproof}
Note that the network service process is characterised by its MGF bound, $ \overline{\mathbb{M}}_{S_{\mathrm{net}}}(\theta,s,t), s >0 $. Furthermore, a  lower bound on the service process  corresponds to an upper bound on $ \overline{\mathbb{M}}_{S_{\mathrm{net}}}(\theta,s,t)$ which is given by \eqref{eqn:lower bound of service capability}. Since $ P_{\sum}=P_{\rm tot} $ by assumption, the optimization problem can be  formulated as
\begin{equation}\label{eqn:lemma3-1}
\mathbf{P}^* = \argmin_{\mathbf{P}:P_{\sum}=P_{\rm tot}} \sum_{\sum\limits_{i=1}^{n+1}\pi_{i}=t-s}\prod_{i=1}^{n+1}\left(\mathbb{E}\left[(1+\gamma_{i})^{-\theta \eta}\right]\right)^{\pi_{i}}.
\end{equation}

To prove the lemma, we use the  \emph{generalized Haber inequality} \cite{belbachir2008multinomial}. That is, for any $ m $ non-negative real numbers $ x_1,x_2,\ldots,x_m $, and for all $ n\geq 0 $, we have
\begin{equation*}
\sum_{i_1+ \dots +i_m = n} \,\,\prod_{k=1}^{m}x_{i}^{i_k}\geq \binom{n+m-1}{m-1}\left(\frac{1}{m}\sum_{k=1}^{m}x_{k}\right)^{n}\,.
\end{equation*}
The above holds with equality  \textit{if and only if} $ x_i=x_j $ for any $ 1\leq i,j \leq m $. Using this result, the minimum in \eqref{eqn:lemma3-1} can be achieved by choosing a power allocation vector $ \mathbf{P}^* $ such that
\begin{equation*}
\mathbb{E}\left[(1+\gamma_{i})^{-\theta \eta}\right] = \mathbb{E}\left[(1+\gamma_{j})^{-\theta \eta}\right]
\end{equation*}
for all $ i, j \in \IM$, which equivalently indicates that the  SINRs are identically distributed  across all $n+1$ hops, i.e., $ \gamma_j \eqd \gamma_i $, for all $i \in \IH$. The resulting MGF service bound  under $ \mathbf{P}^* $ power allocation is then given by
\begin{align}\label{eqn:lemma3-2}
\overline{\mathbb{M}}^*_{S_{\mathrm{net}}}(\theta,s,t)  \le &
 \binom{t \! - \! s \! + \! n}{n} \! \left(\frac{1}{n \! + \! 1}\sum_{i=1}^{n+1}\mathbb{E}\left[(1+\gamma_{i})^{-\theta \eta}\right]\right)^{t-s} \notag \\
{=} & \binom{t   -  s   +   n}{n}\left(\mathbb{E}\left[(1+\hat{\gamma})^{-\theta \eta}\right]\right)^{t-s}\,.
\end{align}
The first step is obtained by applying the  generalized Haber inequality, with \emph{strict equality} under optimal power allocation, to \eqref{eqn:lower bound of service capability}. 
The last step follows since $\gamma_i, \forall i \in \IH$ are i.i.d. under optimal power allocation, and the fact that a random variable is uniquely determined by its MGF, and by letting $ \hat{\gamma} \eqd \gamma_i$. 
%
%
\end{IEEEproof}

Lemma~\ref{lemma:3} shows  that the key enabler of optimal multi-hop network operation is the maintenance of identically distributed channels' SINR across all hops. 
Note that this result applies to arbitrary networks where the service process is characterized by a Shannon-type link model, as long as the individual links are not coupled in terms of interference (only self-interference is considered). 
In contrast, in what follows we present an approach for allocating transmit power\footnote{The power is assumed to be infinitely divisible during the power allocation.}  to achieve identically distributed SINR in a 60~GHz multi-hop network. Given a background noise power of $ N_0 $, we define   $ \lambda_{i}=\frac{P_{i}}{N_{0}} $. Due to the one-to-one correspondence of $\lambda_{i}$ and  $ P_{i} $ and for convenience we opt to work with $\lambda_{i}$ for the  derivations that follows. 

For $ \gamma_i $, $ i\in \mathcal{I}_{\mathcal{H}} $, we use $$ \gamma_i = \kappa \cdot 10^{-0.1\alpha}\omega_i l_i^{-\beta}\cdot 10^{-0.1\xi_i} $$ to rewrite \eqref{eqn: SINR}.
Since the shadowing effects are assumed to be homogeneous, log-normally distributed over all hops, i.e., $ \xi_i\sim \mathcal{N}(0,\sigma^2)$, where the log-normal shadowing variance $ \sigma^2 $ is independent of the transmit power, and since $\omega_i l_i^{-\beta} $  depends on the  power allocated to transmitter $ {i} $, then a power allocation scheme that enables the equality $ \omega_i l_i^{-\beta} = \omega_j l_j^{-\beta} $ for any $ i,j\in \mathcal{I}_{\mathcal{H}} $ is sufficient to ensure identically distributed SINR $\gamma_i, \forall i$.

We follow an iterative approach starting from the last hop of the network, i.e., the $ \left(n+1\right)^{\mathrm{th}} $ hop, and moving backwards. Assume that $  \omega_i l_i^{-\beta} = c$ holds for any $ i\in \mathcal{I}_{\mathcal{H}} $, where $ c $ is a constant. Let $i=n+1$, then according to \eqref{eqn: SINR}, $\omega_{n+1} = \lambda_n$ and therefore, $ \lambda_{n}=c \cdot l_{n+1}^{\beta} $. Similarly, for  $ i=n $, we obtain
\begin{equation*}
\lambda_{n-1}=(1+\mu_{n}\lambda_{n})\cdot c \cdot l_{n}^{\beta} = c \cdot l_{n}^{\beta} + c^2\mu_{n}\left(l_{n}l_{n+1}\right)^{\beta}.
\end{equation*}
Then by recursively applying $ \lambda_{i-1}=\left(1+\mu_{i}\lambda_{i}\right)\cdot c\cdot l_{i}^{\beta} $, for all $i \in \mathcal{I}_{\mathcal{H}}$, and after some manipulation  we obtain
\begin{equation}\label{eqn:generic lambda}
\lambda_i = \sum_{k=1}^{n-i+1}c^{k} \left(\mu_{i+k}^{-1}\prod_{u=1}^{k}{\mu_{i+u} \cdot l_{i+u}^{\beta}}\right)   \triangleq \sum_{k=1}^{n-i+1}\nu_{i,k}\cdot c^k
\end{equation}
for all $0\leq i \leq n$. 
With the constraint $ P_{\sum} \le P_{\rm tot} $, which  corresponds to $ \lambda_{\sum}\triangleq \frac{P_{\sum}}{N_{0}}  \le \frac{P_{\rm tot}}{N_{0}} \triangleq \lambda_{\rm tot}$, we can obtain the value of $ c $ using Lemma \ref{lemma:3} and by solving the equation  $ \lambda_{\sum} = \lambda_{\rm tot} $ subject to
\begin{equation}\label{eqn:lambda_sum for c}
\lambda_{\sum}  = \sum_{i=0}^{n}\left(\sum_{k=1}^{n-i+1}\nu_{i,k}c^k\right) = \sum_{k=1}^{n+1}\left( \sum_{i=0}^{n+1-k}\nu_{i,k}\right)\cdot c^{k},
\end{equation}
where the last equality is obtained by collecting terms containing $ c^{k} $ for $ 1\leq k \leq n+1$. 
Then $ \hat{\gamma} $ can be expressed as
\begin{equation}\label{eqn:averaged sinr}
\hat{\gamma} = \kappa \cdot 10^{-0.1\left(\alpha+\xi\right)} \cdot c.
\end{equation}

\begin{lemma}[Existence]\label{lemma:4}
Given a 60~GHz $(n+1)$-hop  wireless network operating under  transmit power budget $ P_{\rm tot} $, there always exists a unique optimal power allocation $ \mathbf{P}^{*} $ such that $ P_{\sum}=P_{\rm tot} $,   among all feasible $ \mathbf{P}\in \mathbf{\Xi}_{n} $ in terms of maximizing a lower bound on network service process. 
\end{lemma}

\begin{IEEEproof} The proof consists of two parts: (i) the existence and uniqueness of the power allocation scheme given $ P_{\sum} = P_{\rm tot} $, and (ii)  the optimality. 
The notations in derivations below follow the fashion we used previously.

(i) To show that there is exactly one real root $ c $ that meets the constraint on $ \lambda_{\rm tot} $, we first construct $ f(x) $   as follows
\begin{equation}\label{eqn:lemma4-1}
f(x) \triangleq \sum_{k=1}^{n+1}\left( \sum_{i=0}^{n+1-k}\nu_{i,k}\right)x^{k} - \lambda_{\rm tot}.
\end{equation}

Due to the facts that $ f(0) = -\lambda_{\rm tot} < 0 $ and $ f(+\infty)= +\infty $, since $ f(x) $ is continuous, by the \emph{Intermediate Value Theorem}, there is a positive  $ c $ such that $ f(c)= 0 $, which means that equation $ f(x)=0 $ has a root. Furthermore, to prove the uniqueness of the root, we assume equation $ f(x) = 0 $ has at least two positive roots $ x_{1} $ and $ x_{2} $. Suppose that $ x_{1} < x_{2} $  such that $ f(x_{1})=f(x_{2})=0 $. Note that $ f(x) $ is continuous on the closed interval $ [x_{1},x_{2}] $ and differentiable on the open interval $ (x_{1},x_{2}) $, it is easy to find that the three hypotheses of \emph{Rolle's Theorem} are simultaneously satisfied. Thus, there should be a number $ x^{*}\in (x_{1},x_{2}) $ such that $ f'(x^{*})=0 $. However,  
\begin{equation*}
f'\left(x\right) = \sum_{k=1}^{n+1}\left( \sum_{i=0}^{n+1-k}\nu_{i,k}\right)kx^{k-1} > 0
\end{equation*}
holds for any point $ x \in (0,+\infty)  $, which contradicts  the initial assumption and hence we conclude that $ x_{1} = x_{2} $ which proves the uniqueness of the root for the equation $ f(x) = 0 $ and hence the uniqueness of the optimal power control vector.

(ii) To prove the optimality of $\mathbf P^*$, we assume that the optimal solution is obtained with the sum power $ P'_{\sum} $ that relates to the power allocation vector 
$\mathbf P'$, where $ P'_{\sum} < P^{*}_{\sum} = P_{\rm tot}$. In other words, we can obtain $ \hat{\gamma}' $ associated with $\mathbf P'$, such that
\begin{equation}\label{eqn:lemma4-2}
\mathbb{E}\left[(1+\hat{\gamma}')^{-\theta \eta}\right] \leq \mathbb{E}\left[(1+\hat{\gamma}^{*})^{-\theta \eta}\right],
\end{equation}
where $ \hat{\gamma}^{*} $ similarly relates to $\mathbf P^*$. For notational simplicity, by using \eqref{eqn:averaged sinr}, we define
\begin{equation*}
g(c) \triangleq \mathbb{E}\left[(1+\hat{\gamma})^{-\theta \eta}\right]= \mathbb{E}\left[(1+\kappa 10^{-0.1\left(\alpha+\xi\right)}c)^{-\theta \eta}\right].
\end{equation*}

Let us consider all possible power allocation schemes $\mathbf P$ such that  $ P_{\sum} \leq P_{\rm tot} $. It is evident that, the derivative of $ g(c) $ over $ P_{\sum} $ gives
\begin{align}\label{eqn:lemma4-3}
& \frac{d g(c)}{d P_{\sum}} = g'(c) \cdot \frac{dc}{dP_{\sum}} = g'(c) \cdot \left(\frac{dP_{\sum}}{dc}\right)^{-1}  \notag \\
= & -\theta \eta \cdot \mathbb{E}\left[\frac{\kappa 10^{-0.1\left(\alpha+\xi\right)}}{\left(1+\kappa 10^{-0.1\left(\alpha+\xi\right)}c\right)^{1+\theta\eta}}\right]\cdot \frac{N_0}{f'(c)} < 0.
\end{align}
which indicates that $ g(c) $ monotonically decreases in $ P_{\sum} $.  Since $ c $ is uniquely determined by $ P_{\sum} $ as shown above and because of \eqref{eqn:lemma4-3}, we conclude that 
\begin{equation*}
\mathbb{E}\left[(1+\hat{\gamma}')^{-\theta \eta}\right] > \mathbb{E}\left[(1+\hat{\gamma}^{*})^{-\theta \eta}\right],
\end{equation*}
which  contradicts the initial assumption in  \eqref{eqn:lemma4-2}. Thus, the optimal solution is achieved only when $ P_{\sum}= P_{\rm tot} $ is satisfied, which completes the proof.
\end{IEEEproof}

Lemma~\ref{lemma:3} shows that allocating power in such a way that it results in i.i.d. SINRs across the hops is optimal, while 
Lemma~\ref{lemma:4} states that utilizing all the available power for transmission is optimal since it provides the best network performance. The intuition behind these results is that avoiding bottleneck is the best strategy, while utilizing more power for transmission enhances network performance. 
Furthermore, Lemma~\ref{lemma:3} and Lemma~\ref{lemma:4} provide the minimization on the MGF bound of the service process rather than minimizing the actual process. Nevertheless, such minimization results in maximizing the lower bound on network service which in turn enables the computation of better network backlog and end-to-end delay bounds and more efficient resource allocation and network dimensioning when based on the computed results. In light of the above, we have the following result:
\begin{theorem}\label{theorem:power allocation}
Given the total power budget $ P_{\mathrm{tot}} $, i.e., $P_{\sum} \leq  P_{\rm tot}$, and the background noise power $ N_{0} $, for the outdoor $ 60 $ GHz channel described in \eqref{eqn:channel gain}, and let $ x^{*} $ denote the positive solution for the algebraic equation
\begin{equation*}\label{eqn:feasibility condition}
\sum_{k=1}^{n+1}\left( \sum_{i=0}^{n+1-k}\nu_{i,k}\right)x^k = \frac{P_{\mathrm{tot}}}{N_0},
\end{equation*}
with $ \nu_{i,k} $ given by $$ \nu_{i,k} =  \mu_{i+k}^{-1} \cdot \prod_{u=1}^{k}{\mu_{i+u} \cdot l_{i+u}^{\beta}}, $$ 
where $\mu_i$ and $l_i$ are the model parameters defined in Sec.~\ref{Sec: III}, then, there exists a unique optimal power allocation strategy $ \mathbf{P}^{*}\in \mathbf{\Xi}_{n}$,
such that  
\begin{equation*}
P_{i}^{*} = N_0\sum_{k=1}^{n-i+1}\nu_{i,k}\cdot\left(x^{*}\right)^{k}, \; \mbox{ for } i\in\mathcal{I}_{\mathcal{H}} \, .
\end{equation*}
\end{theorem}

\begin{IEEEproof}
Using Lemmas~\ref{lemma:3} and \ref{lemma:4},    the theorem immediately follows by applying the mapping between the transmit power and the SINR, i.e., $ P_{\rm tot}  = \lambda_{\rm tot} N_0$ and $ P_{i}= \lambda_{i} N_0$.
\end{IEEEproof}

\section{Self-interference Impact in 60 GHz Networks}
\label{Sec: V}
To investigate the impacts of self-interference  on network performance, we consider a particular case, where the separation distances between adjacent nodes are assumed to be equal to $ l $, e.g., $ l_{i}=l $ for $ \forall i\in \mathcal{I}_{\mathcal{H}} $, and all relays have an identical self-interference coefficient $ \mu $. Closed-form expressions for the network performance can be obtained under these assumptions which provide more insights to the network operation. 

In the following analysis we assume that the optimal power allocation scheme proposed in Theorem~\ref{theorem:power allocation} is used. Under this power allocation we have, $ c=\omega_{i} l_{i}^{-\beta} $ for  $\forall i\in \mathcal{I}_{\mathcal{H}} $. Note that $c$ in this case is a measure of  the SINR of channels, and hence it directly influences the network performance. Equation \eqref{eqn: SINR} shows that the parameter $ \omega_i $, and therefore  the function $ c $, are functions of $ \mu $, the now variable self-interference coefficient. Therefore, in this section we represent this measure by the function $ c\left(\mu\right)= \omega_{i}(\mu) l_{i}^{-\beta} $. 
Applying \eqref{eqn:lambda_sum for c} under the proposed power allocation scheme as well as the conditions $ P_{\sum}=P_{\rm tot} $ and $\mu_i =\mu $, the optimal $ \lambda_i $ in \eqref{eqn:generic lambda}, denoted by $ \lambda_{i}^{*} $, reduces to
\begin{align}
& \lambda_i^{*} = \mu^{-1}\sum_{k=1}^{n-i+1} \left(c\left(\mu\right)\mu l^{\beta}\right)^{k} \notag \\
= & 
\begin{dcases}
c\left(\mu\right)l^{\beta} \cdot \frac{\left(c\left(\mu\right)\mu l^{\beta}\right)^{n-i+1}-1}{c\left(\mu\right)\mu l^{\beta}-1}, & \mbox{if } c\left(\mu\right)\mu l^{\beta}\neq 1\\
\frac{2\lambda_{\mathrm{tot}}(n-i+1)}{(n+1)(n+2)}, & \mbox{otherwise.}
\end{dcases}
\end{align}

In the above, we used the geometric sum to obtain the first case.
It is easy to find that, the case $ c\left(\mu\right)\mu l^{\beta}= 1 $ corresponds to a particular situation for the self-interference coefficient $ \mu $, i.e.,
$ \mu = \left(2\lambda_{\mathrm{tot}}\right)^{-1} \left(n+1\right)\left(n+2\right) $, which immediately gives the optimal $ c\left(\mu\right) $ as $ c\left(\mu\right)= 2 \lambda_{\mathrm{tot}} \left[\left(n+1\right)\left(n+2\right)l^{\beta}\right] $ by applying $ c\left(\mu\right) = \left(\mu l^{\beta}\right)^{-1} $. It is worth noting that, $ c\left(\mu\right)\mu l^{\beta}= 1 $ here corresponds to a very special case that rarely occurs in realistic scenarios, since $ \mu $, which is only related to the relay implementation, is independent from $ \lambda_{\rm tot} $ and $ n $. In this sense, for most cases $ \mu = \left(2\lambda_{\mathrm{tot}}\right)^{-1} \left(n+1\right)\left(n+2\right) $ is not satisfied. Therefore, from the perspective of generality, our interest mainly focuses on the case $ c\left(\mu\right)\mu l^{\beta}\neq 1 $.

According to the assumption of optimal power allocation that gives $ P_{\sum}=P_{\rm tot} $ according to Lemma \ref{lemma:4},  we have
\begin{align}\label{eqn:polynomial of c}
& \lambda_{\rm tot} = \sum_{i=0}^{n}\lambda_{i}^{*}
= \frac{c(\mu)l^{\beta}}{c(\mu)\mu l^{\beta}-1}\sum_{i=0}^{n}\left(\left(c\left(\mu\right)\mu l^{\beta}\right)^{n-i+1}-1\right) \notag \\
= & \frac{c(\mu)l^{\beta}}{c(\mu)\mu l^{\beta}\! - \! 1} \left(c\left(\mu\right)\mu l^{\beta} \frac{\left(c\left(\mu\right)\mu l^{\beta}\right)^{n+1}-1}{c\left(\mu\right)\mu l^{\beta}-1} \!-\! n \! -\! 1\right),
\end{align}
where we use the change of variables $j=n-i+1$ and then apply the geometric sum formula in the last step. 

For notational simplicity, we set $ t = c(\mu)\mu l^{\beta} $ with $ t\neq 1 $, then \eqref{eqn:polynomial of c} can be written as
\begin{equation*}
\mu\lambda_{\rm tot} = \frac{t}{t-1}\left(t\cdot \frac{t^{n+1}-1}{t-1}-\left(n+1\right)\right),
\end{equation*}
which subsequently gives
\begin{equation}\label{eqn:function of t}
t^{n+3}-\left(n+2+\mu\lambda_{\rm tot}\right)t^2 + \left(n+1+2\mu\lambda_{\rm tot}\right)t = \mu\lambda_{\rm tot}.
\end{equation}

Recovering $ t $ to $ c(\mu)\mu l^{\beta} $ and dividing both sides of \eqref{eqn:function of t} by $ \mu $, in terms of $ c(\mu) $, we immediately obtain
\begin{equation}\label{eqn:root of function of t}
a_1(\mu)c^{n+3}\left(\mu\right)+a_2(\mu)c^{2}\left(\mu\right)+a_3(\mu)c\left(\mu\right) - \lambda_{\mathrm{tot}} = 0,
\end{equation}
where $ a_{i}(\mu) $, $ i=1,2,3 $ are respectively given by
\begin{equation*}
\begin{dcases}
a_1(\mu) = \mu^{n+2}l^{(n+3)\beta},\\
a_2(\mu) = -(n+2+\mu\lambda_{\mathrm{tot}})\mu l^{2\beta},\\
a_3(\mu) = (n+1+2\mu\lambda_{\mathrm{tot}}) l^{\beta}.
\end{dcases}
\end{equation*}

The \emph{Descartes' Sign Rule} \cite{anderson1998descartes} can be applied to determine the number of positive roots of \eqref{eqn:root of function of t}. The rule states that when the terms in a polynomial are ordered according to their  variable exponent, then the number of positive real roots of that polynomial is either the number of sign changes, say $n$, between consecutive non-zero coefficients, or is less than that by an even number, i.e., $n, n-2, n-4,\dots$. 
Clearly, \eqref{eqn:root of function of t} has one or three real positive root(s), if there exists real positive solutions. In addition, we can see that, there are two repeated positive roots $ c_{1}\left(\mu\right)=c_{2}\left(\mu\right)= \left(\mu l^{\beta}\right)^{-1}$. 
By excluding the two repeated roots that violate the condition $ c\left(\mu\right)\mu l^{\beta}\neq 1 $, we are left with one unique positive solution $ c\left(\mu\right)\neq \left(\mu l^{\beta}\right)^{-1} $ for \eqref{eqn:root of function of t}, under optimal  power allocation, which coincides with the claim in Lemma~\ref{lemma:4}.

For arbitrary positive integer $ n $, there is no explicit generic analytical solution of \eqref{eqn:root of function of t} for $ c\left(\mu\right) $. Hence, in order to keep  track of $ c(\mu) $ with respect to $ \mu $, we compute the first derivative over $ \mu $ on both sides of \eqref{eqn:root of function of t}, then we have
\begin{align*}
0 = & a_{1}'(\mu)c^{n+3}(\mu) + \left(n+3\right)a_{1}(\mu)c^{n+2}(\mu)c'(\mu) \\
& + a_{2}'(\mu)c^{2}(\mu) + 2a_{2}(\mu)c(\mu)c'(\mu) \\
& + a_{3}'(\mu)c(\mu) + a_{3}(\mu)c'(\mu),
\end{align*}
which gives
\begin{equation}\label{eqn:derivative of c(mu)}
c'\left(\mu\right)\! = \! -\frac{a_1'(\mu)c^{n+3}\left(\mu\right)+a_2'(\mu)c^{2}\left(\mu\right)+a_3'(\mu)c\left(\mu\right)}
{(n+3)a_1(\mu)c^{n+2}\left(\mu\right)\! +\! 2a_2(\mu)c\left(\mu\right) \! + \! a_3(\mu)}.
\end{equation}

 The  asymptotic behavior of $ c(\mu) $ in the two cases: $ \mu\rightarrow 0 $ and $ n\rightarrow \infty $ is of particular interest and we investigate next. 

\emph{1) For $ \mu \rightarrow 0 $}: 
Note that $ c\left(\mu\right) $ is bounded by $ \frac{\lambda_{\mathrm{tot}}}{(n+1)l^{\beta}} $, since $ \sum_{1}^{n+1} c(\mu)l^\beta = \sum_{1}^{n+1}\omega_{i} \leq \sum_{i=0}^{n} \lambda_{i} = \lambda_{\rm tot}$. Then, \eqref{eqn:derivative of c(mu)} can be expressed by
\begin{equation*}
\lim\limits_{\mu\rightarrow 0}c'\left(\mu\right)=\frac{(n+2)l^{2\beta}c^2\left(\mu\right)-2\lambda_{\mathrm{tot}}l^{\beta} c\left(\mu\right)}{(n+1)l^{\beta}}\leq 0.
\end{equation*}
In the sense of asymptote, we obtain
\begin{equation}\label{eqn:case 1 for small mu}
c(\mu) = \frac{2\lambda_{\rm tot} l^{-\beta}}{n\left(1+\exp\left(\frac{2\lambda_{\rm tot}\mu}{n+1}\right)\right)+2}
\end{equation}
by applying the initial condition $ c(0) = \frac{\lambda_{\mathrm{tot}}}{(n+1)l^{\beta}}$.
 
\emph{2) For $ n\rightarrow \infty $}:
Defining $ D\triangleq \mu\lambda_{\mathrm{tot}} $ and $ t\left(\mu\right)\triangleq c\left(\mu\right)\mu l^{\beta} $, the derivative \eqref{eqn:derivative of c(mu)} can be written as
\begin{equation}\label{eqn:simplifed derivative}
c'\left(\mu\right) = -\frac{c\left(\mu\right)}{\mu}\frac{\left(n^2+(D+3)n+2\right)t\left(\mu\right)-(n+2)D}{\left(n^2+(D+3)n+2+D\right)t\left(\mu\right)-(n+3)D}.
\end{equation}
It is evident that, when $ \lambda_{\rm tot} $ is fixed, $ n\rightarrow \infty $ yields $ n^{-1}D\rightarrow 0 $. Dividing the numerator and denominator of \eqref{eqn:simplifed derivative} by $ n^{2} $, we then have
\begin{equation*}
\lim\limits_{n\rightarrow \infty} c'(\mu) = -\frac{c\left(\mu\right)}{\mu},
\end{equation*}
which immediately gives
\begin{equation}\label{eqn:case 3 for large n}
c(\mu) = K\mu^{-1},
\end{equation}
where $ K>0 $ is a constant scalar.

Following the notation in \eqref{eqn:averaged sinr}, we denote by $ \hat{\gamma}^{*}(\mu) $ the SINR with respect to the optimal power allocation proposed by Theorem~\ref{theorem:power allocation} as a function of self-interference coefficient $ \mu $. Regarding the two asymptotic cases above, we have the following theorem and corollaries to address the relationship between network service capability under optimal power allocation and the self-interference coefficient.

\begin{theorem}\label{theorem:derivative over mu}
Assuming the optimal power allocation in Theorem~\ref{theorem:power allocation}, 
in the two extreme cases discussed above,
the maximized lower bound of network service capability decreases as $ \mu $ grows.
\end{theorem}
\begin{IEEEproof} 
Recalling that the lower bound of network service capability is characterized by $ \mathbb{E}\left[\left(1+\hat{\gamma}^{*}(\mu)\right)^{-\theta \eta}\right] $, we can see that
\begin{equation*}
\begin{split}
& \frac{d}{d\mu}\mathbb{E}\left[\left(1+\hat{\gamma}^{*}(\mu)\right)^{-\theta\eta}\right] = \frac{d}{dc(\mu)}\mathbb{E}\left[\left(1+\hat{\gamma}^{*}(\mu)\right)^{-\theta\eta}\right] c'(\mu)\\
=&  -\theta\eta \cdot \mathbb{E}\left[\frac{\kappa 10^{-0.1\left(\alpha + \xi\right)}}{\left(1+\kappa c(\mu)10^{-0.1\left(\alpha + \xi\right)}\right)^{1+\theta\eta}}\right]\cdot c'(\mu)\\
= &  -\theta\eta \cdot \mathbb{E}\left[\frac{\hat{\gamma}^{*}(\mu)}{\left(1+\hat{\gamma}^{*}\left(\mu\right)\right)^{1+\theta\eta}}\right]\cdot \frac{c'(\mu)}{c(\mu)} > 0,
\end{split}
\end{equation*}
since it has been shown that $ c'(\mu) < 0 $ for asymptotic situations,  e.g., when $ \mu\rightarrow 0 $, which corresponds to case that $ \mu $ is extremely small, or $ n\rightarrow \infty $, which corresponds to the case that $ n $ is sufficiently large. This indicates $ \mathbb{E}\left[\left(1+\hat{\gamma}^{*}(\mu)\right)^{-\theta\eta}\right] $ monotonically increases with $ \mu $, which equivalently shows the degradation of the maximized lower bound of network service capability. Thus, the theorem is concluded.
\end{IEEEproof}

\section{Numerical Results and Discussion}
\label{Sec: VI}
In this section, we provide numerical results for the total backlog and end-to-end delay bounds for multi-hop outdoor $ 60 $ GHz wireless network discussed in Sec.~\ref{Sec: III} and validate them by simulations. Moreover, in the presence of self-interference, the performance of optimal power allocation presented in Sec.~\ref{Sec: IV} and \ref{Sec: V} is demonstrated and further discussed. The common system parameters are listed in Table \ref{tab: system parameter}. Furthermore, we assume  the following regarding the network configuration:
\begin{itemize}
\item Deterministic arrivals with a constant rate $ \rho_a $, and a burst  $ \delta_b=0 $;
\item All relays have identical $ \mu $, and are uniformly deployed along the path from  source to  destination;
\item Sufficiently large (or infinite) buffer size at each relay, i.e., overflow effects are neglected;
\item  A time-slotted system with time intervals of $ 1 $ second are assumed.
\end{itemize}
Under this scenario, we investigate the following:
\begin{enumerate}
\item  We first validate the derived probabilistic delay and backlog upper bounds from Theorem \ref{theorem:upper bound of M}.
\item  Secondly, based on the validated bounds, we investigate the impact of optimal power control, self-interference and relay density on the probabilistic performance of outdoor $ 60 $ GHz multi-hop networks.
\end{enumerate}

For the sake of simplicity, in what follows the analytical bounds for homogeneous scenarios are all illustrated by Theorem~\ref{theorem:homogeneous M}, while heterogeneous counterparts are provided by applying Theorem~\ref{theorem:upper bound of M}. 

\begin{table}[t]
\footnotesize
\renewcommand{\arraystretch}{1.5}
\centering
\caption{System Parameters}
\begin{tabular}{|c|c|c|}
\hline
\textbf{Parameters} & \textbf{Symbol} & \textbf{Value} \\
\hline
\hline
Bandwidth & $ W $ & $ 500 $ MHz\\
\hline
Antenna Gain Scalar & $\kappa$ & $ 70 $ dBi \\
\hline
Power Budget & $ P_{\mathrm{tot}} $ & $ 50 $ Watt  \\
\hline
Noise Power Density& $N_0/W$ & $ -114 $ dBm/MHz\\
\hline
Hop Length & $ l $ & 0.5 km\\
\hline
Path Loss Intercept & $\alpha$ & $ 70 $ \\
\hline
Path Loss Slope & $\beta$ & $ 2.45 $\\
\hline
STD of Shadowing & $ \sigma $ & $ 8 $ dB \\
\hline
\end{tabular}\label{tab: system parameter}
\end{table}

\subsection{Bound Validation}
\begin{figure}
\centering
\includegraphics[width=3in]{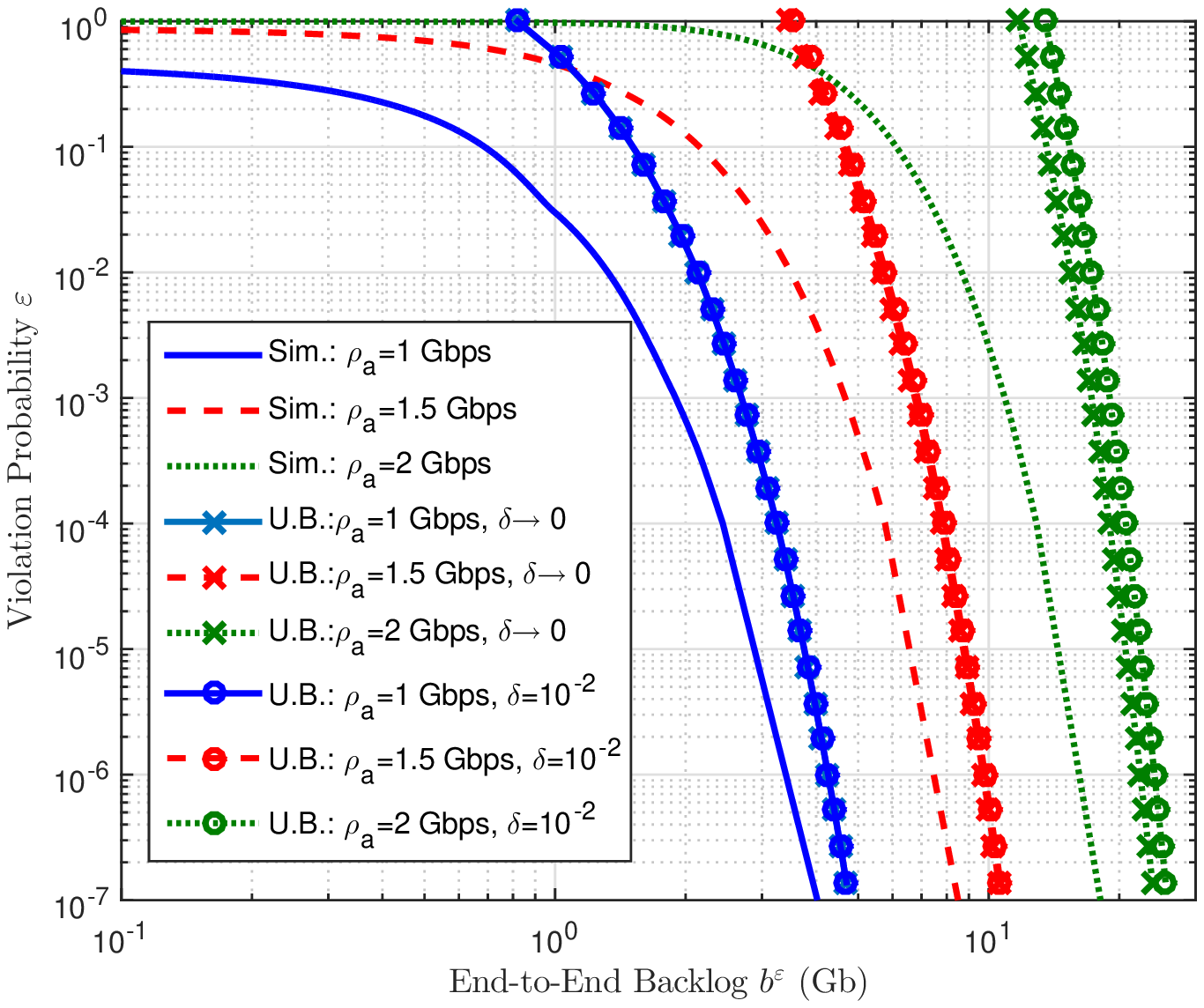}
\caption{Violation probability $ \eps $ v.s. targeted theoretical backlog bounds $ b^{\eps} $, compared to simulations for different $ \rho_a=1 $, $ 1.5 $ and $ 2 $ Gbps, with $ n=10 $, and $ \delta=10^{-2} $ and $ \delta\rightarrow 0 $, respectively.}
\label{fig:fig2a}
\end{figure}

\begin{figure}
\centering
\includegraphics[width=3in]{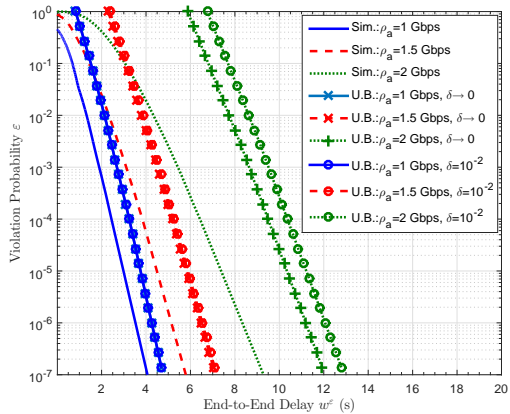}
\caption{Violation probability $ \eps $ v.s. targeted theoretical delay bounds $ w^{\eps} $, compared to simulations for different $ \rho_a=1 $, $ 1.5 $ and $ 2 $ Gbps, with $ n=10 $, and $ \delta=10^{-2} $ and $ \delta\rightarrow 0 $, respectively.}
\label{fig:fig2b}
\end{figure}

We start with considering a tandem 60 Ghz network consisting of $ n=10 $  relays that have identical self-interference coefficient   $ \mu = -80 $ dB. From Table \ref{tab: system parameter} and the power constraint formulated in the form of \eqref{eq: power constraint}, we determine $ \lambda_{\mathrm{tot}}= 134 $ dB. Figs. \ref{fig:fig2a} and \ref{fig:fig2b}  show the  total backlog and the end-to-end delay bounds respectively, compared to  their corresponding simulated values.  Recall that the SINR distributions are identical per hop due to applying the optimal power allocation policy from Theorem \ref{theorem:power allocation}, resulting in $ m=1 $ of Theorem \ref{theorem:upper bound of M}. 

In Fig. \ref{fig:fig2a}, given a violation probability $ \eps $, we observe that the simulated total backlog (the curve without marker) rises as the arrival rate increases from $ 1 $ Gbps to $ 2 $ Gbps due to an increasing utilization of the system. Clearly, the simulated violation probability of backlog asymptotically approaches the analytical upper bound (the curve with marker) as the target end-to-end backlog $ b^{\eps} $ increases. The plot furthermore contains information on the accuracy of the bound provided by Lemma \ref{lemma:inverse moment}, as we compare in the plot the corresponding analytical bounds for a granularity of $ \delta = 10^{-2}$ as well as for letting $ \delta\rightarrow 0 $. We observe that the two curves are quite close together, which confirms also our findings in \cite{2016arXiv160800120Y} that the MGF bound provided by Lemma \ref{lemma:inverse moment} is close to the true MGF of the service process for reasonable granularities $\delta$. The only exception occurs for the end-to-end delay violation probability with a high load of $\rho_a=2 $ Gbps, indicating that the step size might be required to be adapted in certain scenarios. In addition, regarding the end-to-end delay $ w^{\eps} $ with respect to the violation probability $ \eps $ in Fig. \ref{fig:fig2b}, we find that the end-to-end delay is not linearly dependent on the arrival rate, while the delay violation probability bound becomes less accurate as the the utilization approaches the saturation point. Despite this, asymptotically the simulated system behavior and the the bounds show the same slope, concluding our validation of the bounds.

\subsection{Impact of Optimal Power Allocation in 60 GHz Networks}
\begin{figure}
\centering
\includegraphics[width=3in]{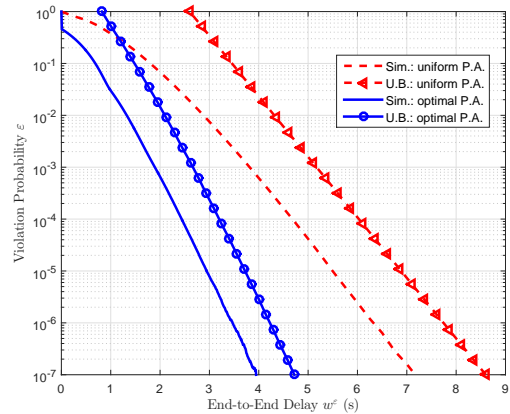}
\caption{Violation probability $ \eps $ v.s. targeted theoretical delay bounds $ w^{\eps} $, compared to simulations for two power allocation strategies, with $ n=10 $, $ \rho_a=1$ Gbps and $ \delta=10^{-2} $.} 
\label{fig:fig3b}
\end{figure}

\begin{figure}
\centering
\includegraphics[width=3in]{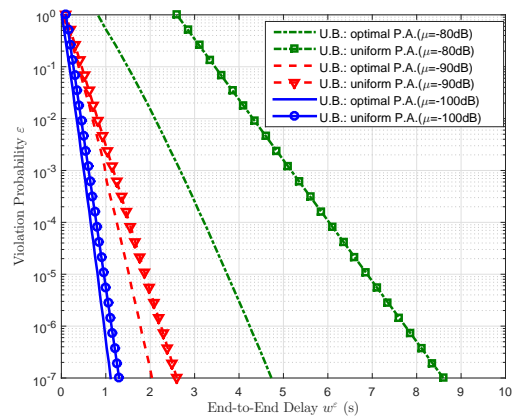}
\caption{Violation probability $ \eps $ v.s. targeted theoretical delay bounds $ w^{\eps} $, for two power allocation strategies, with $ \mu = -80 $ dB, $ -90 $ dB and $ -100 $ dB, respectively, where $ n=10 $, $ \rho_a=1 $ Gbps, and $ \delta=10^{-2} $.} 
\label{fig:fig3d}
\end{figure}

Fig. \ref{fig:fig3b} demonstrates the merit of optimal power allocation, with respect to its impact on the end-to-end delay. We consider a network that consists of $ n=10 $ relays, the self-interference coefficient is $ \mu=-80 $ dB,  and the arrival rate fixed to $ \rho_a=1 $ Gbps.
We use a uniformly allocated powers $ 
\{P_i\}_{i=0}^{n}=\frac{50}{11}\mathrm{W}$ as a baseline for comparison. From the figure it is evident that, the upper bound associated with the optimal power allocation (referring to Theorem~\ref{theorem:homogeneous M} for the homogeneous case) is asymptotically tight, while its counterpart (here applying the result for $ m \geq 2 $ in Theorem~\ref{theorem:upper bound of M}, since the non-optimal power allocation yields the heterogeneity) is not. The slackness of bounds for the heterogeneous scenario comes from producing the binomial coefficient of \eqref{eq:Thm3-1} in Theorem~\ref{theorem:upper bound of M}, where the upper bound is generalized in a simplified and unified manner. In other words, compared to the recursive approach by \cite{petreskarecursive}, the tightness of our proposed method is sacrificed for gaining a lower computational complexity for the heterogeneous cases. Fortunately, the asymptotic tightness can be guaranteed for both homogeneous and heterogeneous scenarios, and this allows us to keep track of realistic performance behaviors. Furthermore, under a sum power constraint,  we can see that, the network performance without the optimal power allocation is subject to severe degradation, in terms of the end-to-end delay. It is evident that the bound performance degradation is also significantly exacerbated as $ \mu $ grows. This obvious deterioration indicates the great importance of adopting  the optimal power allocation by Theorem \ref{theorem:power allocation}, especially when the self-interference is significantly higher than the noise. 

Given different self-interference coefficients, i.e., $ \mu=-100 $ dB, $ -90 $ dB and $ -80 $ dB,  in Fig.~\ref{fig:fig3d},  we furthermore investigate the impact of self-interference on the network performance bound through the optimal power allocation. It is clear that, in the situation $ \mu=-80 $ dB, the optimal power allocation scheme enables a remarkable improvement in terms of performance bounds, while this benefit diminishes as $ \mu $ decreases, as shown from the gaps for $ \mu=-90 $ dB and $ \mu=-100 $ dB. Despite the slackness of upper bound for the heterogeneous cases, we are still able to conclude that optimal power allocation is more important in case of high interference coupling, or low SINR scenarios, in general.

\subsection{Impact of Self-Interference}
\begin{figure}
\centering
\includegraphics[width=3in]{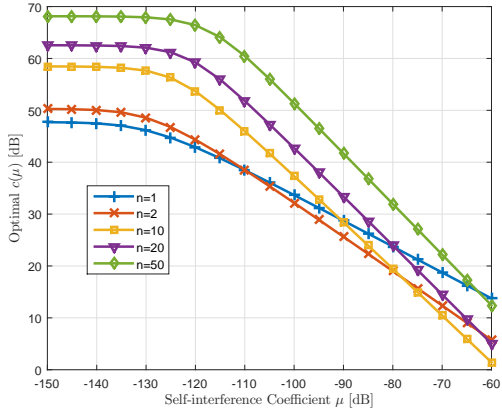}
\caption{ $ c(\mu) $ v.s. $ \mu $ for different relay densities characterized by $ n=1 $, $ 2 $, $ 10 $, $ 20 $ and $ 50 $, respectively, with $ L=5 $ km, $ P_{\mathrm{tot}} =50 $ W.}
\label{fig:fig4}
\end{figure}
In the following, we further investigate the performance of 60 GHz networks operated by optima power allocation while varying the self-interference coupling coefficient $\mu$. We consider the separation distance between source and destination to be fixed to $ L=5 $ km, and an arbitrary number of relays with a sum power constraint is uniformly placed between the source and destination nodes. As we deploy more and more relay, the separation distance decreases as $ l = \frac{L}{n+1} $.

From Lemma~\ref{lemma:3}, we know that the SINR per link, determined by $ c\left(\mu\right) =\omega^{*}\left(\mu\right)l^{-\beta} $, yields the service performance, where $ \omega^{*}\left(\mu\right) $ denotes the optimal $ \omega_{i}\left(\mu\right) $ for $ \forall i\in \mathcal{I}_{\mathcal{H}} $ obtained by the power allocation scheme. We study hence the impact of $ \mu $ on $ c(\mu) $ by this relation, rather than straightforwardly to aim at the probabilistic backlog and delay violation probability bounds. The behavior of $ c(\mu) $ with respect to the varying self-interference coefficient $ \mu $ for different relay densities is shown in Fig. \ref{fig:fig4}. For all curves, there exists a ``waterfall'' shape with regarding $ c(\mu) $ as $ \mu $ increases, dividing the curve into a ``flat'' and ``falling'' stage, respectively. The point from which on this transition happens depends on the node density $n$. Taking the scenario $ n=50 $ as an example, more precisely, when $ \mu $ is below $ \mu_c \approx  -120 $ dB, $ c\left(\mu\right) $ remains flat by increasing $ \mu $. Keeping on increasing $ \mu $, however, $ c\left(\mu\right) $ will encounter a significant decay once $ \mu $ exceeds $ \mu_c $. This behavior relates to the system switching from a noise-limited one to an interference-limited one, where the self-interference becomes the dominant factor restricting the level of $ c\left(\mu\right) $. 
We also observe that, generally speaking, a higher relay density will result in a higher $ c\left(\mu\right) $. However, when the self-interference coefficient is significant and the system being limited by interference, i.e., $ \mu \geq -90 $ dB, a sparser relay deployment is surprisingly able to provide a higher $ c\left(\mu\right) $. From Fig. \ref{fig:fig4}, we summarize that despite improving $ c\left(\mu\right) $ by means of increasing $ n $, the overall performance improves only if $ \mu $ relates not to a strongly interference-limited system. Hence, optimizing the network performance by changing the node density must take the self-interference coefficient into account, since higher relay densities do not always imply higher performance.
\begin{figure}
\centering
\includegraphics[width=3in]{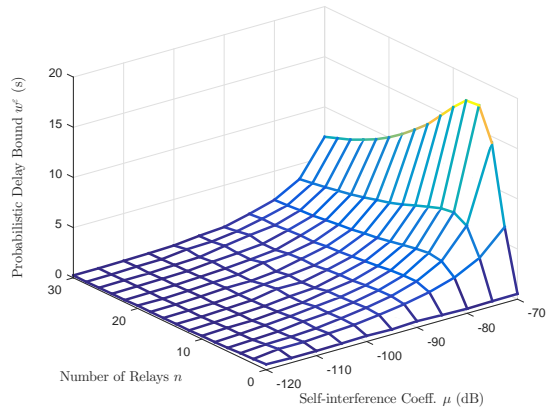}
\caption{Probabilistic delay bound $ w^{\eps} $ v.s. $ n $ and $ \mu $ jointly, with $ \eps = 10^{-6} $.}
\label{fig:fig5}
\end{figure}
This is further demonstrated regarding a varying self-interference coefficient $ \mu $ and a varying number of relays $ n $ with respect to the probabilistic end-to-end delay bound $ w^{\eps} $ in Fig. \ref{fig:fig5}. For more noise-limited systems, the delay bound is barely sensitive to different number of relays (in fact a higher number of relays has a beneficial impact on the delay bound, which is not visible in this plot), while for strongly noise-limited systems either a low or a higher number of relays outperforms relay densities in between significantly. Recall that as the node density increases, the link distances get shorter while the transmit power per relay also decreases. Still, as the results for noise-limited systems demonstrate, the resulting SINR improves as the relay density increases if the self-interference coefficient is small. As the self-interference coefficient increases now, as long as the emitted transmit power per relay creates a \textit{significant} self-interference with the own receiver, the performance degrades. This happens precisely for medium number of relays, while for a larger relay numbers the resulting self-interference per node drops below the noise (asymptotically) leading to a better system performance. 

\section{Conclusions}\label{Sec: VII}
We investigate stochastic performance guarantees, i.e., the probabilistic end-to-end backlog and delay, for multi-hop outdoor $ 60 $ GHz wireless networks with full-duplex buffered relays, by means of MGF-based stochastic network calculus. 
According to specific outdoor propagation features of $ 60 $~GHz radios, a cumulative service process characterization with self-interference is proposed, in terms of the MGF of its channel capacity. 
Based on this characterization, probabilistic upper bounds associated with overall network performance are developed. 
In addition, we propose an optimal power allocation scheme in the presence of self-interference, aiming at enhancing the network performance. 
The analytical framework of this paper supports a broad class of multi-hop networks, in terms of homogeneous and heterogeneous, where the asymptotic tightness of computed upper bounds has been validated. 
Results reveal that, the self-interference coefficient plays a crucial role in improving network performance. 
Another interesting and important finding is that, given the sum power constraint, increasing the relay density does not always improve network performance unless the self-interference coefficient is sufficiently small. 
We believe that approaches developed in this paper will have a variety of applications in designing and optimizing networks for the next generating wireless communications, in terms of performance guarantees and enhancements.


\end{document}